\documentclass[preprint,a4paper,sort&compress]{elsarticle}

\usepackage{lineno,hyperref}
\usepackage{amsmath}
\usepackage{graphicx}% Include figure files
\usepackage{dcolumn}% Align table columns on decimal point
\usepackage{bm}% bold math
\usepackage{multicol}
\usepackage{wasysym}
\usepackage{subcaption}
\usepackage{textcomp}

\usepackage[outdir=./]{epstopdf}

\usepackage{soul,color} % For highlighting.

\journal{the arXiv}

\begin{document}

\begin{frontmatter}

\title{Updated measurement method and uncertainty budget for direct emissivity measurements at UPV/EHU}

\author[address1]{I. Gonz\'alez de Arrieta }

\author[address2]{T. Ech\'aniz\corref{mycorrespondingauthor}}
\cortext[mycorrespondingauthor]{Corresponding author}
\ead{telmo.echaniz@ehu.es}
\author[address2]{R. Fuente}

\author[address1]{J.M. Campillo-Robles}
\author[address1]{J.M. Igartua}

\author[address1]{G.A. L\'opez}

\address[address1]{Applied Physics II, University of the Basque Country UPV/EHU, Leioa, 48940, Spain}
\address[address2]{Applied Mathematics, University of the Basque Country UPV/EHU, Plaza Ingeniero Torres Quevedo 1, Bilbao, 48013, Spain}

\begin{abstract}
This work reports on the upgrades made to the direct emissivity measurement facility of the UPV/EHU. The instrumental improvements consist of, among others, a high-vacuum system and a wider temperature range ($300-1273$ K). Methodological developments include a refined measurement equation with updated parameters and a reworked ISO-compliant uncertainty budget, and a Monte Carlo procedure for accurate calculations of total emissivities from spectral data. These upgrades have been demonstrated and validated in measurements of both metallic and ceramic materials. The results obtained in this work are applicable to similar experimental devices for emissivity measurements in order to report reliable emissivity data.
\end{abstract}


\end{frontmatter}

%\linenumbers

\section{Introduction}

Emissivity is an essential thermophysical property for a number of scientific and industrial applications, but the lack of well-established experimental procedures and reference materials complicates the comparison of data among different research groups. Relatively few attempts at intercomparisons have been pursued, mostly in recent years \cite{LEBARON2019476, Cardenas-Garcia2014, Redgrove1995, Hanssen2016, euramet}. Some ASTM standards are available \cite{ASTM-1, ASTM-2, ASTM-3, ASTM-4, ASTM-5, ASTM-6}, but they do not correspond well to most research devices in operation, which are custom-made by research groups. It is well-known that uncertainty budgets can widely differ among research groups even for standardized techniques, especially for temperature-dependent properties \cite{Rudtsch2005}. In particular, the presence of systematic errors is troublesome and can only be reliably determined by intercomparisons with well-defined uncertainty budgets \cite{PAVESE20091459}. Furthermore, the development of new measurement methods, improved uncertainty calculations and characterization of reference materials have all been identified as key prospects of a recent European Roadmap of Thermophysical Properties \cite{Filtz2015}. Therefore, the development of credible uncertainty budgets and further efforts at standardizing emissivity measurements is a priority.

There are many experimental devices capable of emissivity measurements, which vary in their methodology, calibration procedures, temperature measurement, and heating system, among others. A comprehensive review up to 2015 can be found in Ref.~\cite{Honner:15}, while a remarkable amount of new instruments has been developed more recently \cite{Guo2018, ZHANG20171037, ZHANG2018350, Adibekyan2015, Krenek2015, Urban2017, Zhang:18, Zhang18b, Adibekyan2019, ZHANG2015275, ma12071141, KONG201720, DESOUSAMENESES201596, Giraud2017}. The University of the Basque Country (UPV/EHU) has contributed to this goal with the HAIRL emissometer (High Accuracy InfraRed, Leioa), built from original designs \cite{doi:10.1063/1.2393157} and referenced by some of the newly built instruments. As part of the efforts at improving the metrological quality of this device, it   has recently taken part in a Round Robin test for intercomparison of emissivity measuring devices \cite{LEBARON2019476}. However, some of its features required a significant update in order to be able to cope with new challenges. For example, emissivity measurements for near-room-temperature applications, such as photovoltaic cells, biomaterials, textiles or polymers, require a particularly careful estimation of the uncertainty budget \cite{ZHANG20171037, 0957-0233-12-12-311}. Furthermore, low-emitting materials (such as noble metals) have also been known to require more sensitive treatment of the uncertainty in order to bridge together the often conflicting results from different emissivity measurements setups \cite{WOODS201444, Hameury2018}.

The standard practice for calculating uncertainty budgets is to follow the ISO \emph{Guide to the Expression of Uncertainty in Measurement} (GUM) \cite{gum}. A previous uncertainty budget for the HAIRL device based on such Guide is available in the literature \cite{doi:10.1063/1.3431541}. However, important advances have been made ever since, which motivate an improved reference. Several sources of uncertainty were calculated in a non-GUM-compliant way, while systematic errors were not thoroughly discussed. Besides, only metallic materials were considered, whereas a more general treatment of the uncertainty would be desirable to deal with other materials. This is because the lack of reference materials is a significant obstacle in the improvement of the accuracy of emissometers. ARMCO iron was previously used as a reference material by this laboratory \cite{delCampo2006}, but it is unsuitable for high-temperature measurements due to oxidation and the presence of a structural phase transition. This paper features results for a wider range of materials, including ceramics, which were not covered in the previous uncertainty budget. Their high emissivities also allow them to be measured at temperatures lower than those possible for metals in this setup ($300-373$ K). Finally, no discussion was given in the previous reference to directional or total emissivity measurements and their uncertainties.

In this paper, a review of the updated features and measurement methodology of the HAIRL device, as well as a revised uncertainty budget, are presented. The uncertainty budget is deduced according to the principles expressed in the GUM and it is computed for representative materials covering a range of temperatures, emissivities and emission angles. Calculations of integrated total emissivities and their uncertainty propagation using a Monte Carlo method (as detailed in Supplement 1 to the GUM \cite{gum_montecarlo}), are also derived.

%-------------------------------------------
% MEASUREMENT DEVICE
%-------------------------------------------

\section{Experimental device}
\label{sec:experimental}

\begin{figure*}[t]
\centering
\includegraphics[width = 0.8\linewidth]{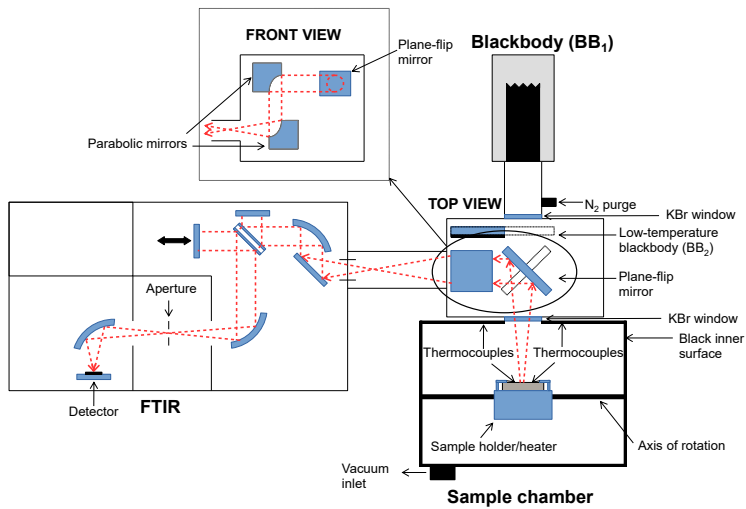} 
\caption{Schematic view of the HAIRL emissometer of the UPV/EHU. The four standalone parts are noted in bold letters, with the optical entrance being shown in both top and front views for better appreciation. The low-temperature blackbody plate is located inside the optical entrance.}
\label{fig:HAIRL}
\end{figure*}

The experimental device used in these measurements is the HAIRL emissometer, which has been in use for more than 14 years and has been constantly updated \cite{doi:10.1063/1.2393157}. It is based on a T-form geometrical configuration described schematically in Fig.~\ref{fig:HAIRL}. It consists of a Fourier-transform infrared spectrometer (FTIR), a vacuum sample chamber, a reference blackbody (Isotech Pegasus R\textsuperscript{\textregistered}) and an optical entrance box that allows switching between the blackbody source and the sample chamber by a rotating plane mirror. The updated technical parameters are listed in Table~\ref{table:parameters}. Note that some of these ranges may not be applicable for all measurements. 

\begin{table}[t]
\centering
\begin{tabular}{ll}
Parameter & Range\\
\hline
Wavelength ($\lambda$, $\mu$m) & $1.43-25$ \\
Viewing angle ($\theta$, $^{\circ}$) & $0-80$ \\
Temperature ($T$, K) & $300-1273$ \\
Pressure ($P$, Pa) & $10^5-5\cdot10^{-3}$\\
Sample atmosphere & Air, Ar, N$_2$+H$_2$ \\
Resolution ($\Delta \sigma$, cm$^{-1}$) & 16 \\
Numerical aperture (NA) & 0.062
\end{tabular}
\caption{List of updated standard parameters for emissivity measurements in the HAIRL device. The spectral resolution is given in wavenumber units by convention ($\sigma = 10^4/\lambda$).}
\label{table:parameters}
\end{table}

The spectrometer is a Bruker IFS 66v/S\textsuperscript{\textregistered} vacuum model. Its optical system consists of conventional KBr optics and a Ge/KBr beamsplitter, with the possibility of using two IR detectors: thermal DLaTGS ($1.43-25$ $\mu$m) and cooled quantum InGaAs ($0.83-2.5$ $\mu$m). The InGaAs detector will not be covered in this work as it has not been as thoroughly tested as the DLaTGS one. The FTIR has the optical entrance system attached directly to its source port, which means it is evacuated to the same degree. This system contains a plane switching mirror to select between the radiance coming from the sample chamber and that coming from the high-temperature blackbody source, as well as two parabolic mirrors to redirect this radiances to the focal point of the FTIR. The additional blackbody source is located inside the optical box, in the form of a steel shutter that has been coated with Nextel 811-21 \textsuperscript{\textregistered} black paint.

The sample chamber consists of a double-wall stainless steel structure, coated with Nextel 811-21 \textsuperscript{\textregistered} black paint to avoid multiple reflections of radiation inside the chamber. A cooling system allows controlling the temperature by running water inside the double wall. The sample holder is located on top of a rotating axis to allow for directional measurements. The vacuum system which controls the sample atmosphere has been upgraded to a turbomolecular pump, which also allows for higher measurement temperatures for metals due to the reduced risk of oxidation. A Zr foil can be used as an oxygen getter for measurements on highly reacting materials, such as titanium or vanadium alloys \cite{ECHANIZ201986}.

The samples are heated from the back by a resistive wire, which was previously embedded inside two steel plates \cite{doi:10.1063/1.2393157}. In this new version, a sample holder with direct radiative contact between sample and wire has been developed. This improves the thermal contact by avoiding thermal gradients across the base plate. The use of a more compact sample holder has allowed for higher measurement temperatures. The regular heating wire is made of Kanthal APM\textsuperscript{\textregistered}, a dispersion-strengthened FeCrAl alloy for use up to 1523 K; whereas a sintered quality Ta heating wire from Plansee (99.95\% purity) can be used for vacuum measurements for temperatures above 1173 K. These improvements allow for sample surface temperatures up to 1273 K (as, for example, in a conductive BN ceramic \cite{2053-1591-3-4-045904}). This new upper temperature limit compares favourably to the previously stated limit of 1050 K in Ref.~\cite{doi:10.1063/1.2393157}.

Typical sample sizes are disks of $\diameter 60$ mm or rectangular samples of $40\times 20$ mm$^2$ in a sample holder of $\diameter 60$ mm. Typical thicknesses range from 0.5 mm to 3 mm, with thicker samples experiencing an excessive axial temperature gradient due to the back-heating configuration. Surface temperature is measured, when possible, by two Type K thermocouples, which are spot-welded to the surface 5 mm away from the center. Each wire is spot-welded independently onto the surface (intrinsic method) to minimize systematic mounting errors or contact spots away from the surface \cite{keltner1983surface}. The wire diameter is 0.2 mm, in order not to excessively disturb the temperature distribution by heat conduction. Alternatively, the thermocouples can be mounted by drilling holes through the material and making mechanical contact at the surface. This method can be described as an intrinsic method with contact resistance \cite{keltner1983surface}. Differences between the two mounting methods will be discussed in Section~\ref{sec:thermocouples}. For ceramic materials, which cannot be welded, an alternative radiometric temperature measurement method is performed, which will be discussed in Section~\ref{sec:Christiansen}.

The high-temperature blackbody employed is a commercial high-emissivity tubular furnace (Isotech Pegasus R\textsuperscript{\textregistered}) with a diameter of $\diameter20$ mm, a depth of 65 mm, and a minimum certified emissivity of 0.995. It is made of graphite and features a diffuse surface and a bottom formed of 120$^{\circ}$ cones to deflect light from the aperture. Its temperature is controlled by a PID controller and measured with a calibrated Type R thermocouple embedded into the bottom of the tube below the textured surface. When measuring its radiance, it can be purged with N$_2$ gas. In contrast, the low-temperature reference is a steel disk painted with Nextel 811-21\textsuperscript{\textregistered} black paint, acting as a greybody. Its temperature is measured by a Type K thermocouple.

%-----------------------------------------------------
% MEASUREMENT METHOD
%-----------------------------------------------------

\section{Measurement and calibration method}
\label{sec:measurement}

\subsection{Previous approach}

The measurement method used at UPV/EHU is based on the \emph{blacksur} method, which has been deemed the most accurate of the simplified direct radiometric methods \cite{Perez-Saez2008}. This method models the sample chamber as a blackbody environment with a much larger area than both the IR window and the sample. The assumption of a large enclosure is the key approximation in this approach, since it simplifies the calculation of the radiation configuration factors and allows considering only a single reflection of a blackbody spectrum. Therefore, the spectral radiance coming out of the chamber window can be described by a simple combination of the self-emitted sample radiation and the blackbody radiation from the enclosure reflected on the sample:

\begin{equation}
L^* = \varepsilon L_s + (1-\varepsilon)L_{sur}
\label{eq:radiancia_efectiva}
\end{equation}
where $L^*$ is the effective spectral radiance, $\varepsilon$ is the spectral emissivity of the sample, $L_s$ is the blackbody spectral radiance of the sample, and $L_{sur}$ is the blackbody spectral radiance of the surroundings.

Assuming a linear instrumental response, typical of thermal detectors like DLaTGS, the following relation between the measured signal and the radiance coming from the sample can be formulated:

\begin{equation}
S_s = RA_sF_{s-det}L^* +S_0
\label{eq:respuesta_lineal}
\end{equation}
where $S_s$ is the measured signal coming from the sample chamber, $R$ stands for the response function of the FTIR, $A_s$ for the sample emission area, $F_{s-det}$ for the radiation geometric factor between the measured spot and the detector, and $S_0$ for the background radiation inside the spectrometer.

The instrument is required to be calibrated with blackbody sources. This procedure is performed by the modified two-temperature method, which has been discussed in the literature and compared satisfactorily to the more common multi-temperature and two-temperature approaches \cite{Gonzalez-Fernandez:10}. In this calibration method, two independent sources of blackbody radiation (a high-temperature source, $bb_1$, and a room temperature one, $bb_2$) are used, which allows for a quicker calibration than by using the same source at two different temperatures. Some of the problems that have been noted for this type of approach \cite{ZHANG20171037} are mostly avoided by measuring the two blackbody sources at very different temperatures ($\Delta T \sim 800 $ K) and keeping one of them near room temperature in order to properly account for the contribution of the background radiation $S_0$ \cite{Gonzalez-Fernandez:10}.

The system of equations required for calibration is thus:
\begin{equation}
S_{bbi} = R A_{bbi} F_{bbi-det}\varepsilon_{bbi}L_{bbi} + S_0 \quad \textrm{for } i=1,2
\label{eq:respuesta_referencias}
\end{equation}

Since both radiance references are located along the same optical path, $A_{bb1} F_{bb1-det} = A_{bb2} F_{bb2-det}$ was assumed. If the observed areas and configuration factors corresponding to the sample and reference measurements also coincide, then the definition of the $R$ factor can incorporate them through $RA_sF_{s-det} = RA_{bb1}F_{bb1-det} \equiv R^*$. Finally, the previously employed measurement equation for the \emph{blacksur} method is obtained by combining Eqs.~\ref{eq:radiancia_efectiva} to \ref{eq:respuesta_referencias} \cite{doi:10.1063/1.3431541}:

\begin{equation}
\varepsilon = \frac{S_s - S_0 - R^* L_{sur}}{R^*(L_s-L_{sur})}
\label{eq:blacksur}
\end{equation}

On a final note, Eq.~\ref{eq:blacksur} is applicable only when there are no additional (parasitic) radiance signals different from $S_0$ that need to be filtered out. It has been proved that this is only possible for off-normal emission angles, as additional contributions can arise if a reflecting sample is located perfectly parallel to the KBr window and the detector \cite{ECHANIZ201722}. Therefore, measurements reported as normal emissivities actually correspond to directional measurements at an incidence angle of $10^{\circ}$.

%-----

\subsection{New approach}

Eq.~\ref{eq:blacksur} has been recently modified in order to account for additional measurement parameters and improve its reliability. Firstly, the possibility of an anisotropic response in the form factors of both sides of the optical entrance is taken into consideration by introducing an anisotropic response function $a(\lambda)$ to account for the chromatic aberration due to different focal points at the left and right sides of the optical entrance. This is assumed to stem from a combination of residual misalignments in the optical entrance and FTIR. This results in the configuration factors for the sample and the references not being equal ($F_{s-det} = aF_{bb-det}$, where $a$ is a measurable optical path difference factor). Secondly, the emissivities of the blackbodies were not considered in the previous equation. The emissivity of the commercial high-temperature blackbody was assumed equal to 1, whereas the emissivity of the low-temperature blackbody source was the only parameter considered. Thirdly, the calibration parameters $R$ and $S_0$ are correlated because they are the solutions of a system of two equations (Eq.~\ref{eq:respuesta_referencias}). Therefore, their uncertainties cannot be considered separately, so it is more effective not to include them in the measurement equation. By considering explicitly the reference data (radiances and emissivities) in Eq.~\ref{eq:blacksur}, a more reliable expression for the \emph{blacksur} measurement method is obtained:

\begin{equation}
\varepsilon = \frac{\frac{(S_s/a-S_{bb1})\cdot(\varepsilon_{bb1}L_{bb1}-\varepsilon_{bb2}L_{bb2})}{S_{bb1}-S_{bb2}}+\varepsilon_{bb1}L_{bb1}-L_{sur}}{L_s-L_{sur}}
\label{eq:telmo}
\end{equation}

This measurement equation can be simplified by defining a ratio quantity $Q$ that includes all radiance measurements:

\begin{equation}
Q = \frac{S_s/a-S_{bb1}}{S_{bb1}-S_{bb2}}
\label{eq:q}
\end{equation}

It must be noted that all measured radiances $S_i$ are signed functions, which can be positive or negative depending on the wavelength and temperature of the source. In particular, it is well established that the $S_{bb2}$ contribution is negative at room temperature, as the detector emits more radiation when aiming at a cold surface than the one that it receives from it \cite{Gonzalez-Fernandez:10}. The ratio parameter $Q$ now includes all relevant radiance measurements onto one single function, leaving all the others solely as functions of non-radiometric parameters.

The final measurement equation is:

\begin{equation}
\varepsilon = \frac{Q\cdot(\varepsilon_{bb1}L_{bb1}-\varepsilon_{bb2}L_{bb2})+\varepsilon_{bb1}L_{bb1}-L_{sur}}{L_s-L_{sur}}
\label{eq:measurement_final}
\end{equation}

This equation bears strong resemblance to one used at Physikalisch-Technische Bundesanstalt (PTB) for their in-air and vacuum emissivity measuring apparatuses \cite{Monte_2010, Adibekyan2015}. This formulation allows for separation of the sources of uncertainty in three uncorrelated sets of parameters: signals ($Q$), temperatures ($T_s$, $T_{sur}$, $T_{bb1}$, $T_{bb2}$), and emissivity of the references ($\varepsilon_{bb1}$, $\varepsilon_{bb2}$). The resulting parameters are advantageous for uncertainty calculations, unlike those of the previous calibration ($R$ and $S_0$). It should be reminded, however, that there is a correlation between the emissivity values at different wavelengths (i.e., $r(\varepsilon(\lambda_i),\varepsilon(\lambda_j))\neq 0$), which will be important in the calculation of total emissivities in Section~\ref{sec:monte_carlo}.

%
%	UNCERTAINTIES
%
\section{Sources of uncertainty in spectral measurements}
\label{sec:sources}

\begin{table*}[hpbt]
\caption{Uncertainty sources for emissivity measurements using the HAIRL device and their respective subsources. Type A corresponds to sources evaluated using statistical methods and Type B to those evaluated by other means \cite{bipm2012international}. The standard uncertainties of quantities described by $t$-distributions are calculated following Section 6.4.9 of Ref.~\cite{gum_montecarlo}. N.S.: not significant.}
\begin{center}
\begin{tabular}{llll}
Source of uncertainty & Symbol & Type & Distribution \\
\hline
1) Signal ratio  & $Q$ & &  \\
\hspace{0.5cm} Sample signal repeatability & $S_s$ & A & Gaussian \\
\hspace{0.5cm} High-$T$ blackbody signal repeatability & $S_{bb1}$ & A & Gaussian \\
\hspace{0.5cm} Low-$T$ blackbody signal repeatability & $S_{bb2}$ & A & Gaussian \\
\hspace{0.5cm} FTIR non-linearity & & A & N.S. \\
\hspace{0.5cm} Size-of-source effect & & B & N.S. \\
\hspace{0.5cm} FTIR drift & & B & N.S. \\
\hspace{0.5cm} Optical path difference & $a$ & A & Gaussian \\
2) Sample surface temperature & $T_s$  &  &  \\
\hspace{0.5cm} Metals & & & \\
\hspace{1cm} Repeatability and inhomogeneity & & A & $t$-distribution \\
\hspace{1cm} Thermocouple (K) accuracy & & B & Rectangular \\
\hspace{0.5cm} Ceramics & & B & Rectangular \\
3) Surroundings temperature & $T_{sur}$ &  &  \\
\hspace{0.5cm} Repeatability and inhomogeneity & & A & $t$-distribution \\
\hspace{0.5cm} Thermocouple (K) accuracy &  & B & Rectangular \\
\hspace{0.5cm} Emissivity of the surroundings & $\varepsilon_{sur}$ & B & N.S. \\
4) Blackbody references & & & \\
\hspace{0.5cm} High-$T$ blackbody temperature & $T_{bb1}$ &  &  \\
\hspace{1cm} Repeatability & & A & Gaussian \\
\hspace{1cm} Inhomogeneity& & B & Gaussian \\
\hspace{1cm} Thermocouple (R) accuracy & & B & Gaussian \\
\hspace{0.5cm} Low-$T$ blackbody temperature & $T_{bb2}$ &  &  \\
\hspace{1cm} Repeatability & & A & Gaussian \\
\hspace{1cm} Thermocouple (K) accuracy & & B & Rectangular \\
\hspace{0.5cm} High-$T$ blackbody emissivity & $\varepsilon_{bb1}$ & B & Rectangular \\
\hspace{0.5cm} Nextel 811-21 emissivity & $\varepsilon_{bb2}$ & B & Gaussian 
\end{tabular}
\end{center}
\label{table:budget}
\end{table*}

The uncertainty budget of the \emph{blacksur} method was first introduced in Ref. \cite{doi:10.1063/1.3431541} and will be revised in this section, according to the new measurement equation (Eq.~\ref{eq:measurement_final}), and the new estimations of uncertainty for the base input quantities. It is based on the linearized GUM framework \cite{gum}, considering only first-order uncorrelated terms. Two types of uncertainty components are considered in this framework: Type A stands for components evaluated by statistical methods, whereas Type B stands for components evaluated using other methods (e.g., calibration reports) \cite{bipm2012international}. A summary of the findings of this Section can be found in the uncertainty budget shown in Table~\ref{table:budget}. The sensitivity factors are obtained by partial derivation of Eq.~\ref{eq:measurement_final}:

\begin{equation}
\frac{\partial\varepsilon}{\partial Q} = \frac{\varepsilon_{bb1}L_{bb1}-\varepsilon_{bb2}L_{bb2}}{L_s-L_{sur}}
\end{equation}
\begin{equation}
\frac{\partial\varepsilon}{\partial L_s} = \frac{L_{sur}-\varepsilon_{bb1}L_{bb1}-Q\cdot(\varepsilon_{bb1}L_{bb1}-\varepsilon_{bb2}L_{bb2})}{(L_s-L_{sur})^2}
\end{equation}
\begin{equation}
 \frac{\partial\varepsilon}{\partial L_{bb1}} = \frac{\varepsilon_{bb1}(1+Q)}{L_s-L_{sur}}
\end{equation}
\begin{equation}
 \frac{\partial\varepsilon}{\partial L_{bb2}} = -\frac{\varepsilon_{bb2}Q}{L_s-L_{sur}}
\end{equation}
\begin{equation}
\frac{\partial\varepsilon}{\partial L_{sur}} = \frac{\varepsilon_{bb1}L_{bb1}+Q(\varepsilon_{bb1}L_{bb1}-\varepsilon_{bb2}L_{bb2})-L_{s}}{(L_s-L_{sur})^2}
\end{equation}
\begin{equation}
\frac{\partial\varepsilon}{\partial \varepsilon_{bb1}} = \frac{L_{bb1}(1+Q)}{L_s-L_{sur}}
\end{equation}
\begin{equation}
\frac{\partial\varepsilon}{\partial \varepsilon_{bb2}} = -\frac{QL_{bb2}}{L_s-L_{sur}}
\end{equation}

%\vspace{0.4cm}

It should also be borne in mind that, in the case of the $L_i$ radiances, the real uncertainties correspond to the temperature measurements $T_i$, and so it is necessary to introduce derivatives of Planck's law \cite{howell2015thermal}:

\begin{equation}
L(\lambda,T) = \frac{2\pi C_1}{\lambda^5 (e^{C_2/\lambda T}-1)}
\label{eq:planck}
\end{equation}
where $C_1=hc^2$ and $C_2=hc/k_B$ are the first and second radiation constants, formed from fundamental constants.

The partial derivative with temperature then follows as:

\begin{equation}
\frac{\partial L_i}{\partial T_i} = \frac{2\pi C_1 C_2 e^{C_2/\lambda T_i}}{\lambda^6 T_i^2 (e^{C_2/\lambda T_i}-1)^2} = \frac{L_i}{\lambda T_i^2}\frac{C_2e^{C_2/\lambda T_i}}{(e^{C_2/\lambda T_i}-1)}
\end{equation}

Therefore, the final expression for the combined standard uncertainty $u_c$ is:

%\begin{equation}
%\begin{split}
%u_c^2(\varepsilon) &= \bigg ( \frac{\partial\varepsilon}{\partial Q} \bigg )^2 u^2 (Q) + \bigg ( \frac{\partial\varepsilon}{\partial L_s}\frac{\partial L_s}{\partial T_s} \bigg )^2 u^2 (T_s)\\
% &+ \bigg ( \frac{\partial\varepsilon}{\partial L_{bb1}}\frac{\partial L_{bb1}}{\partial T_{bb1}} \bigg )^2 u^2 (T_{bb1}) + \bigg ( \frac{\partial\varepsilon}{\partial L_{bb2}}\frac{\partial L_{bb2}}{\partial T_{bb2}} \bigg )^2 u^2 (T_{bb2}) \\
% &+ \bigg (\frac{\partial\varepsilon}{\partial L_{sur}}\frac{\partial L_{sur}}{\partial T_{sur}} \bigg )^2 u^2 (T_{sur}) + \bigg ( \frac{\partial\varepsilon}{\partial \varepsilon_{bb1}} \bigg )^2 u^2 (\varepsilon_{bb1}) \\
% &+ \bigg ( \frac{\partial\varepsilon}{\partial \varepsilon_{bb2}} \bigg )^2 u^2 (\varepsilon_{bb2})
%\end{split}
%\end{equation}

\begin{equation}
\begin{split}
u_c^2(\varepsilon) &= \bigg ( \frac{\partial\varepsilon}{\partial Q} \bigg )^2 u^2 (Q) + \sum_{i=1}^4\bigg ( \frac{\partial\varepsilon}{\partial L_i}\frac{\partial L_i}{\partial T_i} \bigg )^2 u^2 (T_i)\\
 &+ \sum_{i=1}^2\bigg ( \frac{\partial\varepsilon}{\partial \varepsilon_{bbi}} \bigg )^2 u^2 (\varepsilon_{bbi})
\label{eq:final_uncertainty}
\end{split}
\end{equation}

\subsection{Signal ratio measurement}

Repeatability in the measurement of the emitted radiances is one of the clearest sources of uncertainty in emissivity measurements, particularly for low-emitting samples or at low temperatures and short wavelengths. This is evaluated as a Type A uncertainty, that is, arising from a statistical analysis of Eq.~\ref{eq:q} (where all three radiance measurements are described by uncorrelated Gaussian distributions):

\begin{equation}
u^2(Q) = \sum_{i=1}^3 \bigg (\frac{\partial Q}{\partial S_i} \bigg )^2 u^2(S_i) + \bigg (\frac{\partial Q}{\partial a}\bigg )^2 u^2(a)
\label{eq:std_q}
\end{equation}

\begin{figure}[hptb]
\centering
\includegraphics[width=\linewidth]{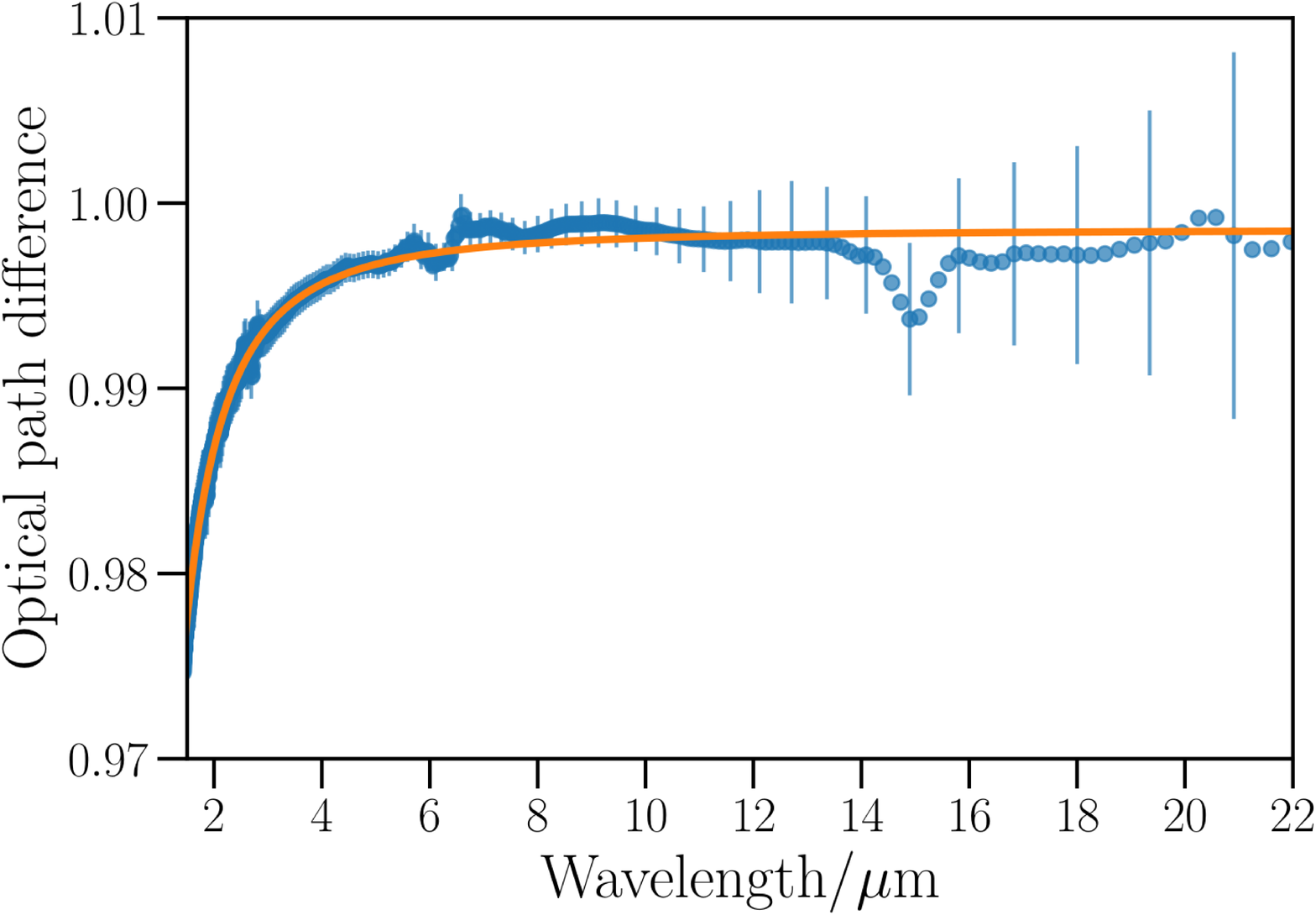}
\caption{Experimentally determined optical path difference factor ($a$) and fitting. Error bars correspond to the standard uncertainty ($k=1$). Note that the features around 6 and 15 $\mu$m correspond to residual absorption by water and CO$_2$. Data around 4 $\mu$m have been interpolated to avoid the strongest CO$_2$ absorption mode.}
\label{fig:anisotropy}
\end{figure}

The uncertainty corresponding to the optical path difference factor $a$ is routinely checked, especially during maintenance of the FTIR, by measuring the radiance emitted by an infrared source located at each side of the optical entrance. The result of the latest calibration is shown in Fig.~\ref{fig:anisotropy}. The standard uncertainty is calculated as a Type A uncertainty of 30 scans for each side of the entrance (sample and references). The uncertainty was larger at both ends of the spectrum due to lower radiances taking place at such wavelengths. 

As this calibration procedure needs to be done in air for practical reasons, the resulting spectra shows fluctuations, especially around the atmospheric absorption bands. Because of this, the factor which is used in the calculations has been fitted to the simplest functional form that can replicate it (for smoothing purposes):

\begin{equation}
a = \frac{S_R}{S_L} = m_1+\frac{m_2}{\lambda^{2}}
\end{equation}
where $S_R$ and $S_L$ stand for the measured calibration signals from the right (sample) and left (blackbody) compartments. %The resulting parameters were found to be $m_1=0.99859\pm0.00003$ and $m_2=-0.0472\pm0.0001$. %The standard uncertainty of the fitting in the last significant digit is shown in parantheses.

The optical paths followed by both sides of the interferometer diverge increasingly with decreasing wavelength. Such spectral dependence is coherent with the presence of non-ideal surfaces, such as the optical entrance box in our setup \cite{chamberlain1979principles}. The fitted curve and the standard uncertainty of the measurement are introduced into Eq.~\ref{eq:std_q}. Practical reasons do not allow for very frequent recalibrations of this parameter, but the drift with time is sufficiently slow so as to be negligible in this uncertainty budget. %Nevertheless, it should be kept in mind that it is a potential source of systematic errors, which can even lead to non-physical values of the emissivity for very high- or low-emitting samples.

Other sources can contribute to the radiance factor uncertainty. Non-linearity of the FTIR spectrometer with a thermal DLaTGS detector has been checked by a multi-temperature calibration procedure \cite{ZHANG20171037} and found to be indistinguisable from the standard uncertainty of the procedure. In the case of the size-of-source effect, the size of the blackbody aperture is $\diameter 20$ mm and that of the usual samples is greater than $\diameter 20$ mm, while the measuring spot corresponding to the standard aperture size is estimated as $\diameter 3.6$ mm. It has been shown by other researchers that this effect can generally be neglected in emissivity measurements if both radiation source apertures are much larger than the measuring spot size \cite{0957-0233-12-12-311, Cardenas14}. Finally, interferometer drift is minimized by controlling the room temperature. This variation has been checked for experimentally in a previous study and deemed negligible within the allowed 1-2 K variation range \cite{Gonzalez-Fernandez:10}.

%-------------

\subsection{Sample surface temperature measurement}

When considering the standard uncertainty of the temperature measurements in this experimental setting, two cases will be discussed: contact measurements with thermocouples for metals and non-contact pyrometric measurements using the Christiansen wavelength for ceramics \cite{Rousseau2005}. The former was first introduced in a different manner in the previous uncertainty budget \cite{doi:10.1063/1.3431541}, whereas the latter has not yet been examined in our laboratory.

\subsubsection{Metals (thermocouples)}
\label{sec:thermocouples}

There are two  main sources of uncertainty when measuring temperatures with thermocouples that will be discussed in this section. The first one, of Type A, concerns the temporal and spatial thermal inhomogeneity, as measured directly with the thermocouple probes. The second source, of Type B, corresponds to the intrinsic accuracy of the sensors (Type K thermocouples). 

The first source is estimated based on the readings of the two thermocouples, located symmetrically across the measuring spot. Start and end temperatures are recorded for each sensor, leading to four temperature datapoints per measurement. Thus, the inhomogeneities may be described as a $t$-distribution with a number of degrees of freedom $\nu=4-1$. Spatial and temporal inhomogeneities are treated equally, as experience shows them to be correlated. Taking a Bayesian approach (as described in Section 6.4.9 of the Suplement 1 to the GUM \cite{gum_montecarlo}), it is possible to define an enlarged standard uncertainty for measurements characterized by the $t$-distribution. The use of Eq.~\ref{eq:bayesian} simplifies the calculation of the uncertainty by avoiding the need to take the number of degrees of freedom into account \cite{gum_montecarlo}:

\begin{equation}
u(x) = \bigg ( \frac{n-1}{n-3}\bigg )^{1/2} \frac{s}{\sqrt{n}}
\label{eq:bayesian}
\end{equation}
where $n$ is the number of measurements (in this case, $n=4$), and $s=[\sum(x_i-\overline x)^2/(n-1)]^{1/2}$ is the standard deviation.

Regarding the second source, the accuracy of a standard Type K thermocouple is given by the ANSI C96.1 standard as a limit of error of 0.75\% of the temperature in $^{\circ}$C or 2.2 K, whichever is greater \cite{sandia}. In the past uncertainty budget published by this group, these values were introduced into the budget as the standard uncertainty of a Gaussian distribution \cite{doi:10.1063/1.3431541}. However, the limits of error define the boundaries of the allowed uncertainty for a thermocouple conforming to the required composition \cite{sandia}. Therefore, the standard uncertainty of this measurement does not correspond to these limits, but must be estimated from them. Some groups have considered these limits of error as equivalent to a near-total 99.7\% certainty with a coverage factor of $k=3$  \cite{sandia}, while others have taken these limits as $k=2$ \cite{Teodorescu2008}. In this work, a rectangular distribution function has been considered, as recommended by the GUM when only an upper and a lower bound are available \cite{gum}. By assuming this distribution, the standard uncertainty related with this source becomes $1/\sqrt{3}$ of either limit of error, whichever is greater:

\begin{equation}
u(T_{\textrm{TC}}) =\frac{1}{\sqrt{3}} \cdot \bigg\{\begin{array}{lr} 0.0075(T-273.15) \\ 2.2 \text{ K} \end{array}
\label{eq:thermocouple}
\end{equation}

Therefore, the resulting uncertainty is formulated by combining the two sources of uncertainty, assuming that the accuracies of both thermocouples are correlated (as they come from the same batch of material). It is thus given by:

\begin{equation}
u^2(T) = u^2(T_{\textrm{TC}}) + \bigg ( \frac{n-1}{n-3}\bigg ) \frac{s^2(t_i)}{n}
\label{eq:final_temperature}
\end{equation}
where $u(T_{\textrm{TC}})$ is given by Eq.~\ref{eq:thermocouple}.

It should be borne in mind that the uncertainty of the thermocouple response may not always correspond exactly to that defined by Eq.~\ref{eq:thermocouple}. The main problem observed with Type K thermocouples is the potential effect of systematic deviations arising from problems such as magnetic inhomogeneities and selective oxidation of the Chromel leg. Both issues can lead to faulty surface temperature measurements. An extensive literature on the biases and problems associated with Type K thermocouples has been produced \cite{MANGANO1993171, Kollie75, Pavlasek2015, Webster2017, Webster2017_2}. Therefore, an upgrade to a Type N thermocouple is projected for the HAIRL to avoid these systematic errors.

\begin{figure}[hptb]
\centering
\includegraphics[width=\linewidth]{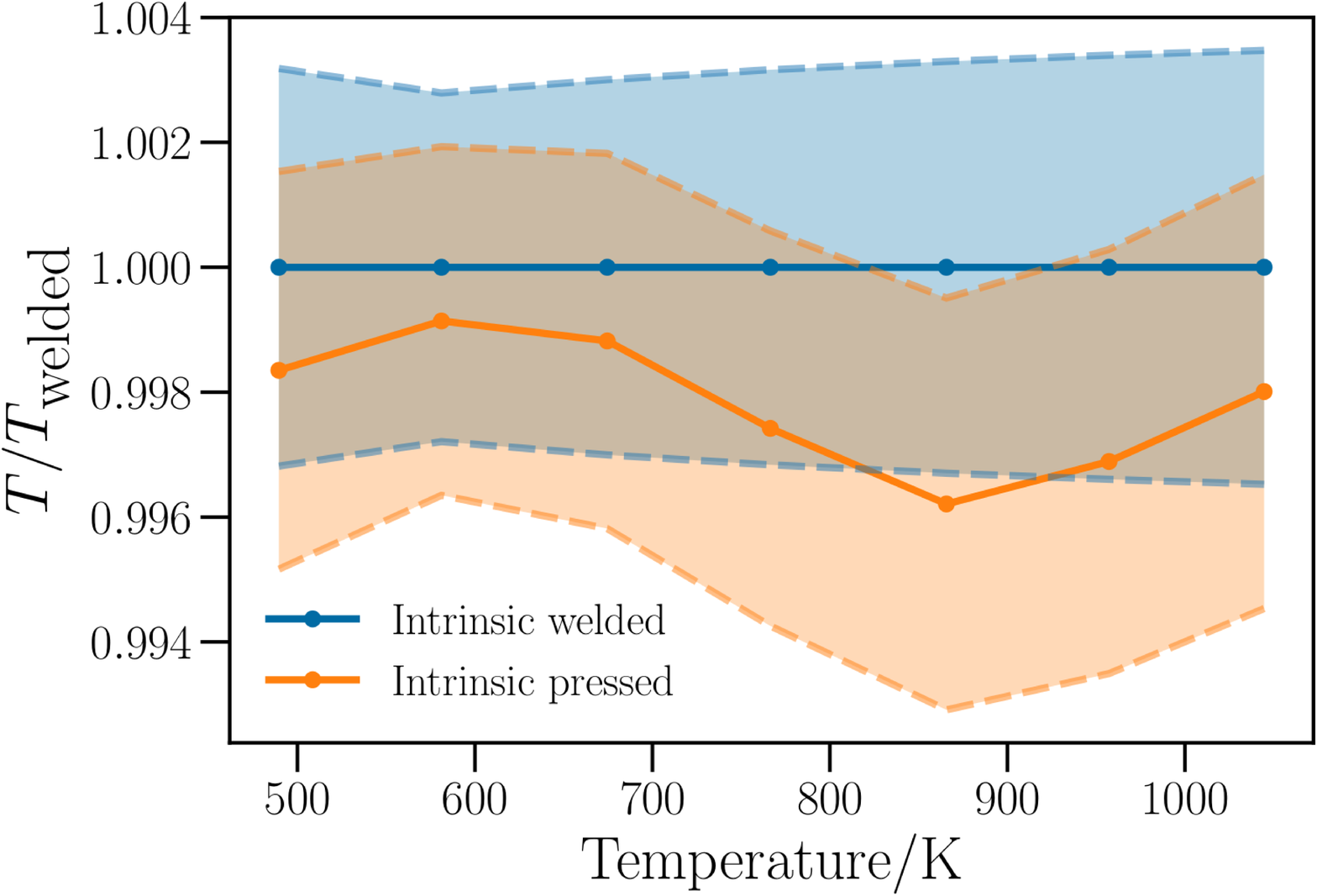}
\caption{Study of the systematic error in temperature measurement by thermocouples drilled and inserted perpendicularly to the surface, when compared to those spot-welded onto the surface for an ARMCO iron disk in Ar atmosphere. Shaded areas lines correspond to the relative standard uncertainties of both methods at each temperature.}
\label{fig:termopares}
\end{figure}

As a final note, the equivalence of both methods for mounting thermocouples mentioned in Section~\ref{sec:experimental} has been tested for an ARMCO iron sample in an Ar atmosphere for the usual stabilization time (20 minutes). The differences between the intrinsic welded and intrinsic pressed methods can be seen in Fig.~\ref{fig:termopares}. Both curves agree within less than the standard uncertainty for most temperatures, with no clear temperature dependence. The magnitude of this effect is reduced in vacuum due to the absence of convection effects. This systematic error arises from an imperfect thermal contact in the mechanical method, but it is not considered in the final uncertainty budget, as it is difficult to estimate for each material and application. Both methods lead to the same temperatures in the infinite-time limit \cite{keltner1983surface}, which means that the mechanical method is a suitable alternative given larger stabilization times compared to the welded method. Accordingly, systematic uncertainties arising from mounting errors are considered as the main source of uncertainty in high-temperature transient measurements \cite{sandia}. This can contribute to an additional uncertainty source when using this method at the highest temperatures possible with this setup ($\sim 1273$ K).

\subsubsection{Ceramics (non-contact measurement)}
\label{sec:Christiansen}

Temperature measurements in ceramic materials are difficult to accomplish with contact temperature sensors because of their low thermal and electrical conductivities. Therefore, a non-contact temperature measurement method is desired, although the effect of the emissivity complicates this procedures. Fortunately, some ionic compounds have a well-defined wavelength at which the emissivity is close to 1 and is only very weakly dependent on temperature. This wavelength is known as the Christiansen point and has been used as a reliable method for measuring the surface temperatures of heteropolar ceramic materials \cite{Rousseau2005}. Temperature measurements using the Christiansen wavelength are very useful because they are performed at the same time as the sample radiance measurements and largely avoid systematic errors due to thermal gradients, since the temperature is directly measured on the measuring spot itself. However, no widely recognized estimation of its uncertainty is available at the moment.

An error of $10$ K in the determination of the melting point of single-crystal $\alpha$-Al$_2$O$_3$ has been determined in a similar setup during emissivity measurements, which allows an estimation of $u(T_{Chris}) \sim 0.5\%$ at very high temperatures ($T\geq 2000$ K) \cite{DESOUSAMENESES201596}. Similar relative uncertainties have been reported as extremes for FTIR-based temperature measurements \cite{Dufour:98}, as well as in comparisons between the Christiansen wavelength method and a more complete one based on thermal flux balances and knowledge of the thermophysical properties of polycrystalline $\alpha$-Al$_2$O$_3$ \cite{ZHANG2018350}. %The Christiansen wavelength method yielded systematically higher values than the iterative method, which was deemed to be the best reference, but always less than 0.7\%, and if the highest error is not considered, less than 0.4\%.
Even in cases where the Christiansen-wavelength emissivity changes with temperature, these variations are consistent with such an upper boundary (i.e., the relative change of the spectral emissivity around that wavelength is below 1.5\%, which is within the standard uncertainty of the emissivity when the 0.5\% error in temperature is assumed) \cite{Cagran2007}. Therefore, the relative standard uncertainty in the sample surface temperature associated with the Christiansen method is evaluated in this work as a Type B uncertainty modelled by a rectangular distribution with an upper bound of 0.5\%.

It should be noted that model-based temperature measurement methods can in fact be less accurate than radiometric ones under certain circumstances, as they strongly depend on the accuracy of numerous input parameters \cite{hanssen2007use}. These results serve to prove the usefulness of the Christiansen wavelength as an alternative to more complex measurement methods, with the added advantage of measuring the local temperature of the central spot, without effects introduced by radial gradients. Such a comparison to a thermal-flux based method has not been performed in our laboratory because the heating system employed does not allow a straightforward heat conduction modelling.
%In order to check the potential evolution of such measurements, we check similar studies in SiC with an external non-contact temperature measurement method \cite{Cagran2007}. The results show that, even though the emissivity of SiC does shift upwards and to longer wavelengths around the Christiansen wavelength, the relative change of the spectral emissivity around that wavelength is below 1.5\%, which is within the standard uncertainty of the emissivity when the 0.5\% error in temperature is assumed.

\subsection{Surroundings temperature measurement}

The treatment of the uncertainties arising from the chamber walls emission is similar to that of the sample temperature. It is treated as a blackbody ($\varepsilon=1$) around room temperature, with its temperature measured by Type K thermocouples in two spots in the chamber walls which are symmetric with respect to the optical path of the measurement. As the temperature is less than 300 K (due to the water cooling system), the standard uncertainty of the thermocouples (Eq.~\ref{eq:thermocouple}) is taken as $2.2/\sqrt{3}$ K. The inhomogeneity of the enclosure temperature is calculated as that of the sample temperature; i.e., as a Type A uncertainty corresponding to measurements of two thermocouples at two different times. A Bayesian Type A uncertainty for small number of observations is again formulated, resulting in the same expression as Eq.~\ref{eq:final_temperature}.

Due to the use of a water cooling system, the average enclosure temperature is $\simeq285$ K, instead of 298 K (the value assumed in the previous uncertainty budget \cite{doi:10.1063/1.3431541}). The effect of the reduction in enclosure temperature is negligible for high sample temperatures, but makes precise measurements possible below 330 K. This is due to the fact that the greatest limitation of the \emph{blacksur} method is that the emissivity diverges when the temperature of the sample and the surroundings are very close (see the denominator in Eq.~\ref{eq:blacksur}). Finally, another possible source of uncertainty for this radiance is the assumption that the sample surroundings behave as a blackbody, whereas the emissivity of the coating employed has been measured to be closer to 0.97 \cite{Adibekyan2017}. However, it has been proved that the \emph{blacksur} method differs less than $0.05\%$ from the most accurate method (multi-reflections) when the emissivity of the chamber walls is greater than 0.95, provided that the sample chamber area is much larger than the spot size \cite{Perez-Saez2008}. We therefore neglect the influence of this source of uncertainty.

\subsection{Blackbody references}

The measurement method requires radiance measurements of a high-temperature blackbody and a room-tempe\-rature Nextel 811-21\textsuperscript\textregistered coating. The uncertainty of the measured signals has already been taken into account in the calculation of the $Q$ ratio. The emissivities of the sources are $\varepsilon_{bb1} > 0.995$ for the conventional blackbody source (as specified by the manufacturer), and a spectral average of $\varepsilon_{bb2} = 0.97$ (with a standard uncertainty of $0.01$) for the Nextel 811-21\textsuperscript\textregistered  coating, as calculated from literature data \cite{Adibekyan2017}. Not having any additional information on the uncertainty of the emissivity of the high-temperature blackbody, we consider only the uncertainty given by the calibration certificate (issued by the manufacturer and traceable to the National Physical Laboratory in the UK), modelled as a rectangular probability distribution function (PDF) with limits $0.995<\varepsilon_{bb1}<1$. Both uncertainties were neglected in the previous uncertainty budget \cite{doi:10.1063/1.3431541}, but will be considered in this updated version. In the case of the reference temperatures, their standard uncertainties are taken as the combination of a 0.3 K inhomogeneity in the blackbody temperature distribution and 1 K (expanded uncertainty of 2 K, $k=2$) for the calibration certificate of the high-temperature blackbody (measured with a Type R thermocouple), and $2.2/\sqrt{3}$ K for the low-temperature blackbody (Type K thermocouple, Eq.~\ref{eq:thermocouple}).

Finally, it should be noted that all two-temperature methods are susceptible of systematic errors. Two-tempera\-ture methods have been claimed to incur in significant errors in the determination of the internal radiation sources \cite{ZHANG20171037}. This can be crucial for certain measurements with low signals (low-emitting materials or room-temperature measurements). In the case of this instrument, its calibration accuracy has been checked to be satisfactory given large enough ($\geq$ 800 K) thermal differences between both blackbody references, with one of them required to stay at room temperature to properly account for the aforementioned internal radiation sources \cite{Gonzalez-Fernandez:10}. This contrasts with the approach followed by other authors, who choose blackbody references which have the most similar radiation temperature to the sample \cite{Adibekyan2015}. In both cases, two-temperature methods are considered to be suitable given a linear detector, such as DLaTGS, and sufficiently good control of the blackbody temperatures and the spectrometer drift.

\subsection{Combined standard uncertainty of representative materials}

\begin{figure}[hptb]
\centering
\includegraphics[width=\linewidth]{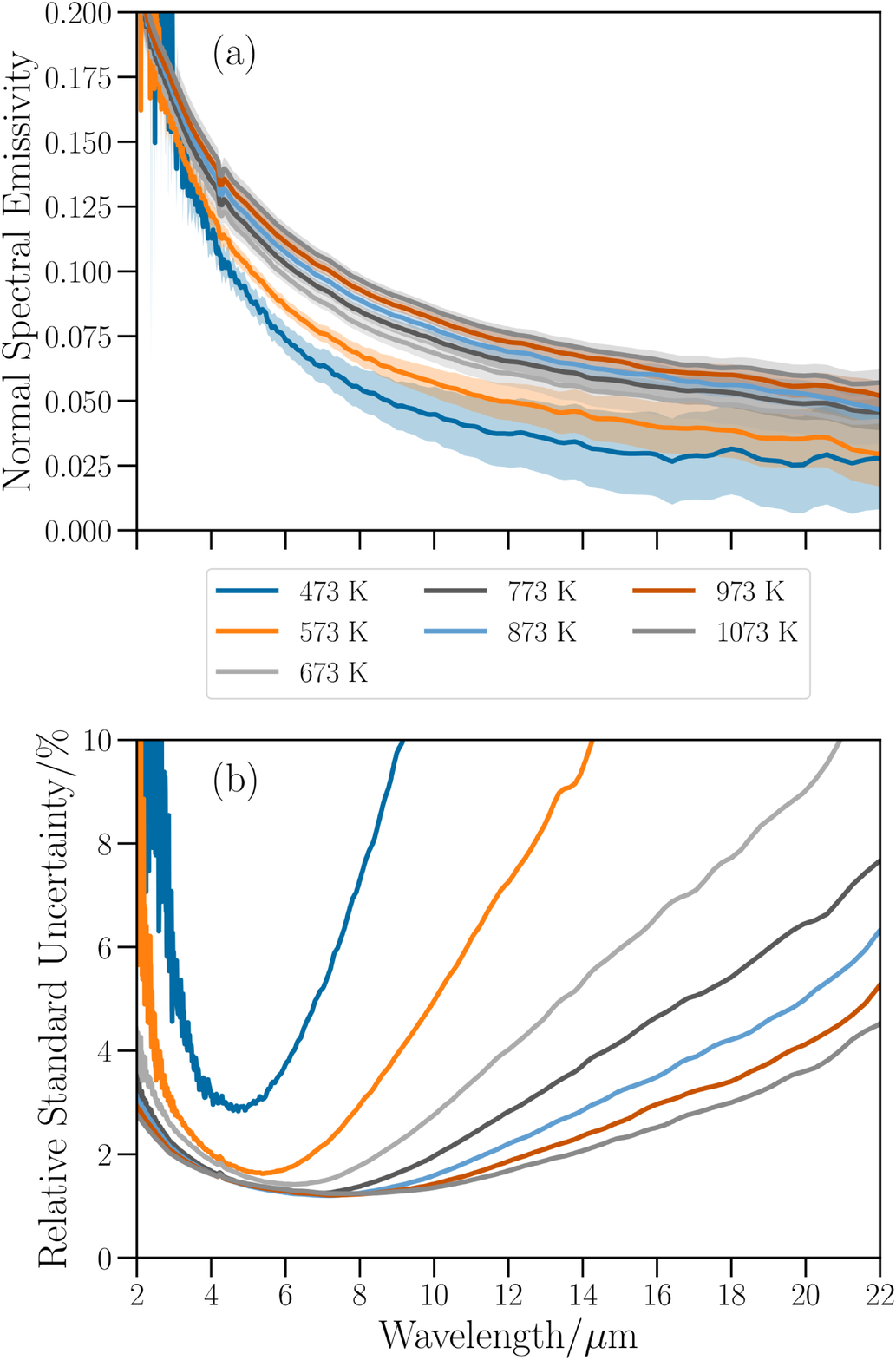}
\caption{(a) Normal spectral emissivity of Ni as a function of temperature. Shaded regions correspond to expanded uncertainties ($k=2$). (b) Relative standard uncertainties ($k=1$) of the normal spectral emissivity measurements of Ni.}
\label{fig:Ni_sombra}
\end{figure}

The uncertainty budget described above for spectral emissivity measurements is applied in this subsection to two materials, a metal and a ceramic. Nickel was chosen as the metallic example based on previous experience with this material \cite{GONZALEZDEARRIETA2019270}. A sputtering target synthesized using the Mond process with a diameter of 50 mm and a thickness of 3.2 mm was used. Its nominal purity was $>99.99\%$, with 15 ppm Fe and 10 ppm S as the main impurities. It was mechanically polished with alumina powder to an average surface roughness of $R_a=0.03$ $\mu$m and a root-mean-square value of $R_q=0.04$ $\mu$m, as measured with a mechanical profilometer (Mitutoyo SJ201). Its emissivity and standard uncertainty are shown in Fig.~\ref{fig:Ni_sombra}a.
The uncertainty tendencies can be more easily appreciated as relative uncertainties in Fig.~\ref{fig:Ni_sombra}b. It can be seen that the uncertainty is very large at 473 K for all wavelengths, and quickly drops with temperature. Regarding its wavelength dependence, it tends to increase with wavelength for all temperatures, except at short wavelengths and low temperatures, where it can also reach large values. It is interesting to note that the previous uncertainty budget reported generally greater uncertainties at shorter rather than larger wavelengths \cite{doi:10.1063/1.3431541}. 

In the case of ceramic materials, a 0.43 mm-thick single crystal of Al$_2$O$_3$ has been chosen. It is oriented perpendicular to the $c$ axis with a 0.2$^{\circ}$ misorientation angle. It was epi-polished on one side to a roughness of $R_a<0.3$ nm. It was synthesized by the Czochralski method, with a purity of $>99.99\%$. The temperature was measured assuming an emissivity of 1 at the Christiansen wavelength, as checked using specular reflectance measurements. Due to its semi-transparency at shorter wavelengths, it was measured using a low-emissivity iron substrate in order to reduce the spurious radiation from the  highly-emitting heating plate in the back. The method for compensating this effect is described in Ref.~\cite{Jeon:10}:

\begin{equation}
\varepsilon = \frac{\varepsilon^*-\tau\varepsilon_{sub}}{1+(1-\varepsilon_{sub})\tau}
\label{eq:semitransparency}
\end{equation}
where $\varepsilon^*$ is the apparent emissivity without semitransparency corrections, $\tau$ is the transmittance of the sample, and $\varepsilon_{sub}$ is the emissivity of the substrate.

Thus, two additional sources need to be included in the uncertainty calculations and all partial derivatives evaluated in Section~\ref{sec:sources} need to take into account an additional weight of $(1+(1-\varepsilon_{sub})\tau)^{-1}$. The transmittance is evaluated as a Type B uncertainty from extreme values taken from literature data \cite{Lee2011}, whereas the emissivity of ARMCO iron and its combined standard uncertainty have been evaluated in our laboratory.

\begin{figure}[hptb]
\centering
\includegraphics[width=\linewidth]{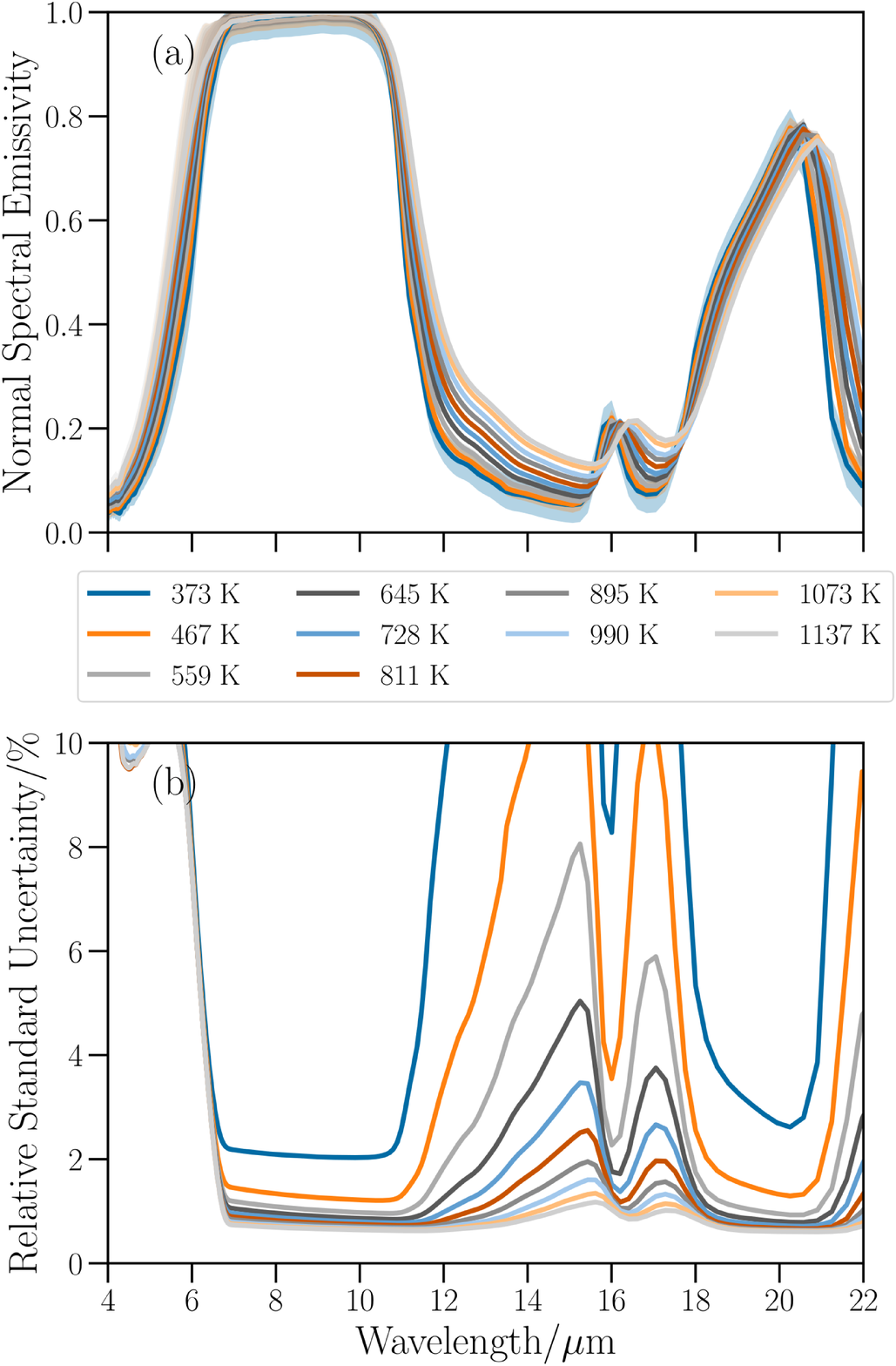}
\caption{(a) Normal spectral emissivity of sapphire (single-crystal Al$_2$O$_3$) as a function of temperature. Shaded regions correspond to expanded uncertainties ($k=2$). (b) Relative standard uncertainties ($k=1$) of the normal spectral emissivity measurements of sapphire Al$_2$O$_3$.}
\label{fig:Al2O3_sombra}
\end{figure}

The emissivity measurements of sapphire and their relative standard uncertainties can be seen in Figs.~\ref{fig:Al2O3_sombra}a and \ref{fig:Al2O3_sombra}b. It is clear that the relative uncertainties at higher temperatures are much lower than those of nickel, mainly due to the higher emissivity values (except in the semi-transparent range below 7 $\mu$m). This shows that, for most of the mid-infrared range, the lowest relative standard uncertainties achievable using the HAIRL device are around $1\%$. Nevertheless, the relative uncertainty at low temperatures is significant, particularly in the low-emissivity regions at long wavelengths. These results on sapphire demonstrate the capability of the apparatus to measure the emissivity of ceramic materials, as well as to deal accurate emissivity values at relatively low temperatures, even for emissivity values of 0.2 or below. 

As can be seen in Figure~\ref{fig:Al2O3_sombra}, the standard uncertainty for ceramic materials is low for most wavelengths. This stems from the greater signal-to-noise ratio of these high-emissivity spectral regions, as well as the reduced contamination from low-temperature radiances (surroundings, detector). Nevertheless, these measurements were performed down to 373 K, almost 200 K less than the lowest temperature considered in the previous uncertainty budget \cite{doi:10.1063/1.3431541}. Emissivity measurements at such low temperatures are easier for high-emitting ceramics than for metals. The temperature can be reduced even further in the case of highly emitting materials, given that the calibration procedure is accurate enough and that the temperature of the vacuum chamber and the sample are sufficiently different. The latter is the main drawback of the \emph{blacksur} method, which has a divergence in the denominator (Eq.~\ref{eq:blacksur}) when those two temperatures become equal \cite{Perez-Saez2008}.

\begin{figure}[hptb]
\centering
\includegraphics[width=\linewidth]{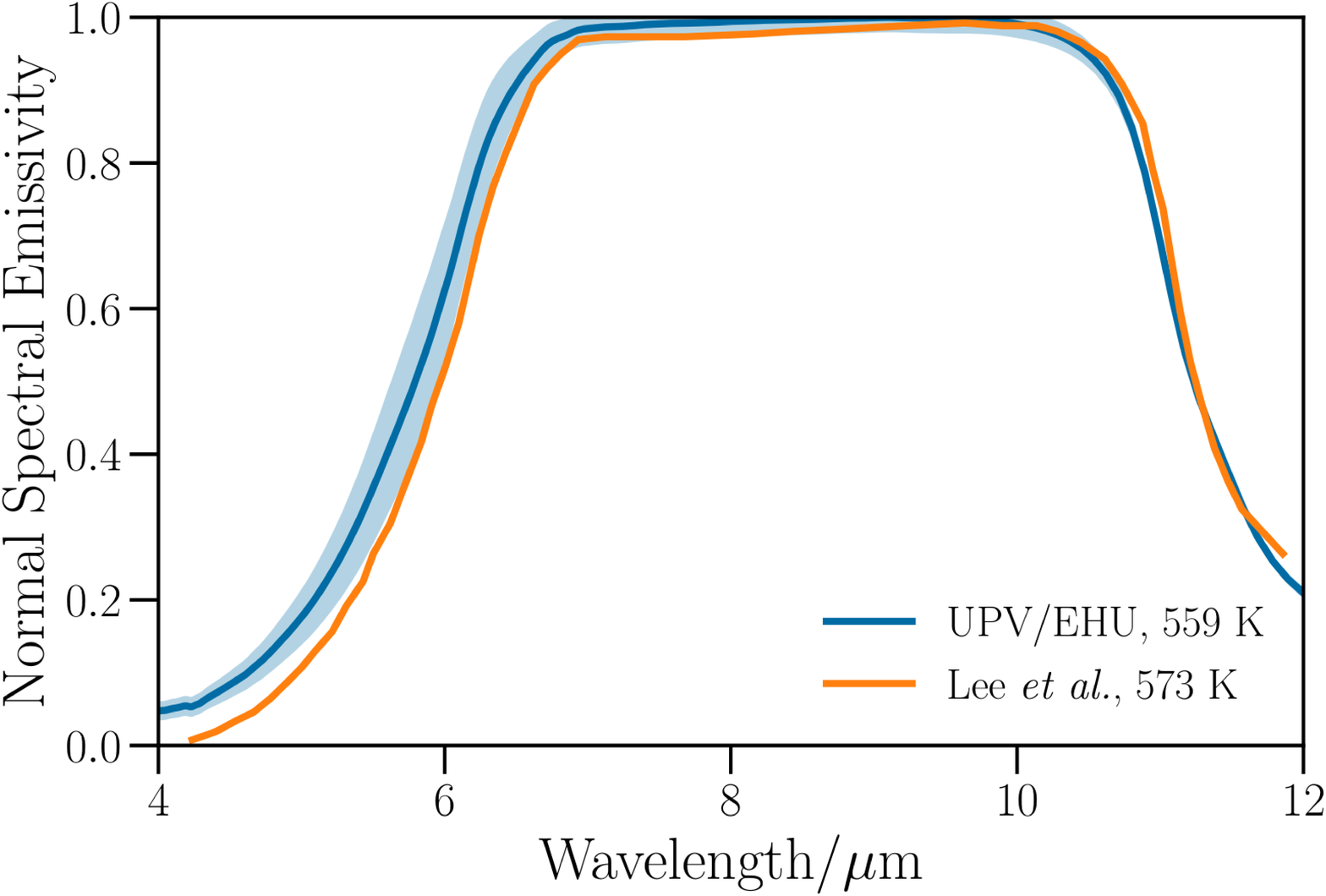}
\caption{Normal spectral emissivity of sapphire Al$_2$O$_3$ measured with the HAIRL device at UPV/EHU at 559 K, compared to literature data at a similar temperature \cite{Lee2011}. Shaded regions correspond to expanded uncertainties ($k=2$).}
\label{fig:comparison}
\end{figure}

In order to check the validity of the applied method and the calculated uncertainties, a comparison to literature data from Ref.~\cite{Lee2011} has been made in Fig.~\ref{fig:comparison}. There is quantitative agreement between both curves in certain wavelength ranges, although some discrepancies may be noted. In particular, the low-wavelength limit of the data measured with the HAIRL emissometer in the fully transparent region is higher than the one reported in the reference. A possible explanation may be due to a mismatch between the temperatures of the sample and the substrate, as Eq.~\ref{eq:semitransparency} implicitly assumes no thermal gradients. Future work is to be carried out on this topic.

\begin{figure}[hptb]
\centering
\includegraphics[width=\linewidth]{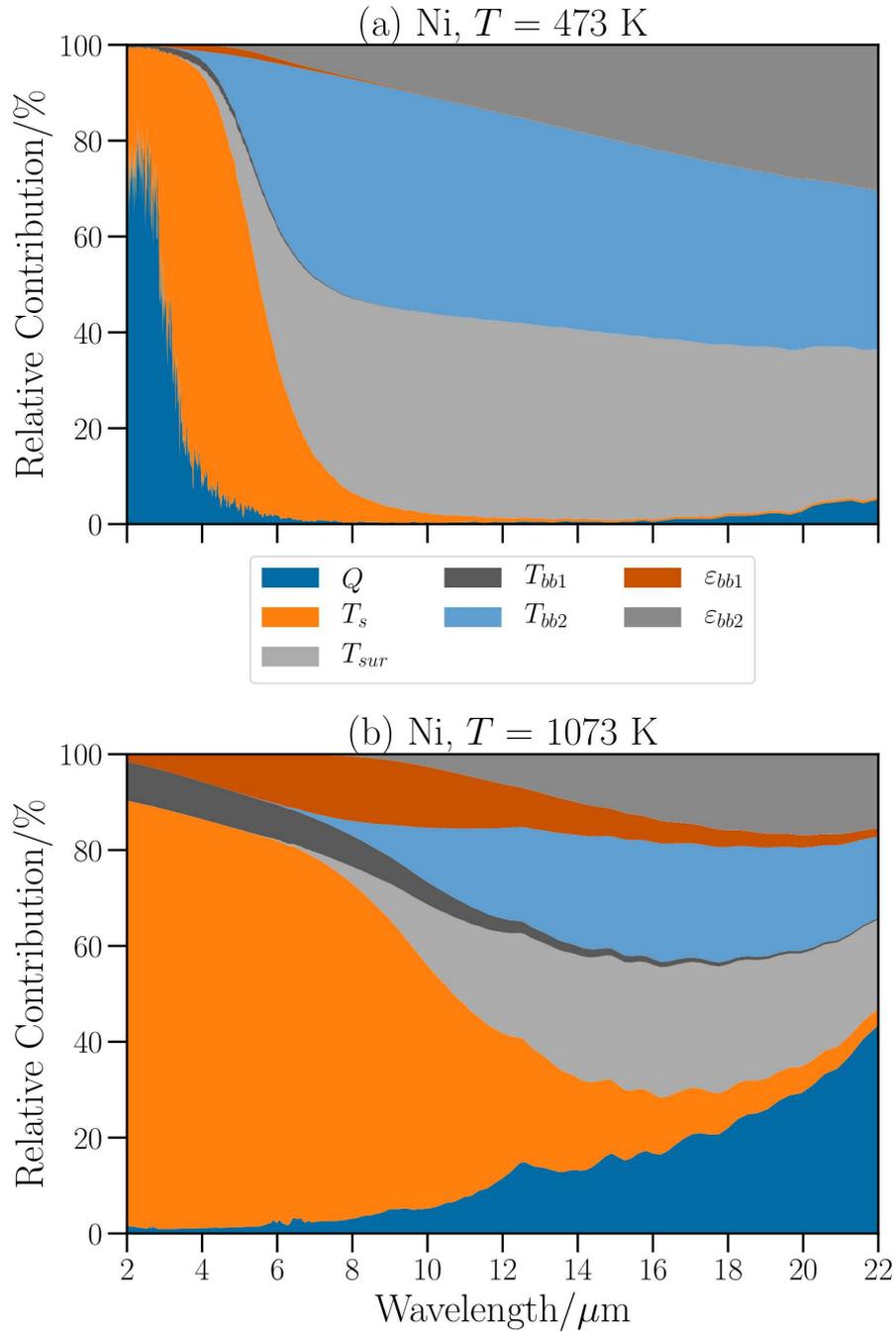}
\caption{Relative contribution of each term of Eq.~\ref{eq:final_uncertainty} to the total variance for Ni at the lowest (a) and highest (b) temperatures measured.}
\label{fig:Ni_relativas}
\end{figure}

\begin{figure}[hptb]
\centering
\includegraphics[width=\linewidth]{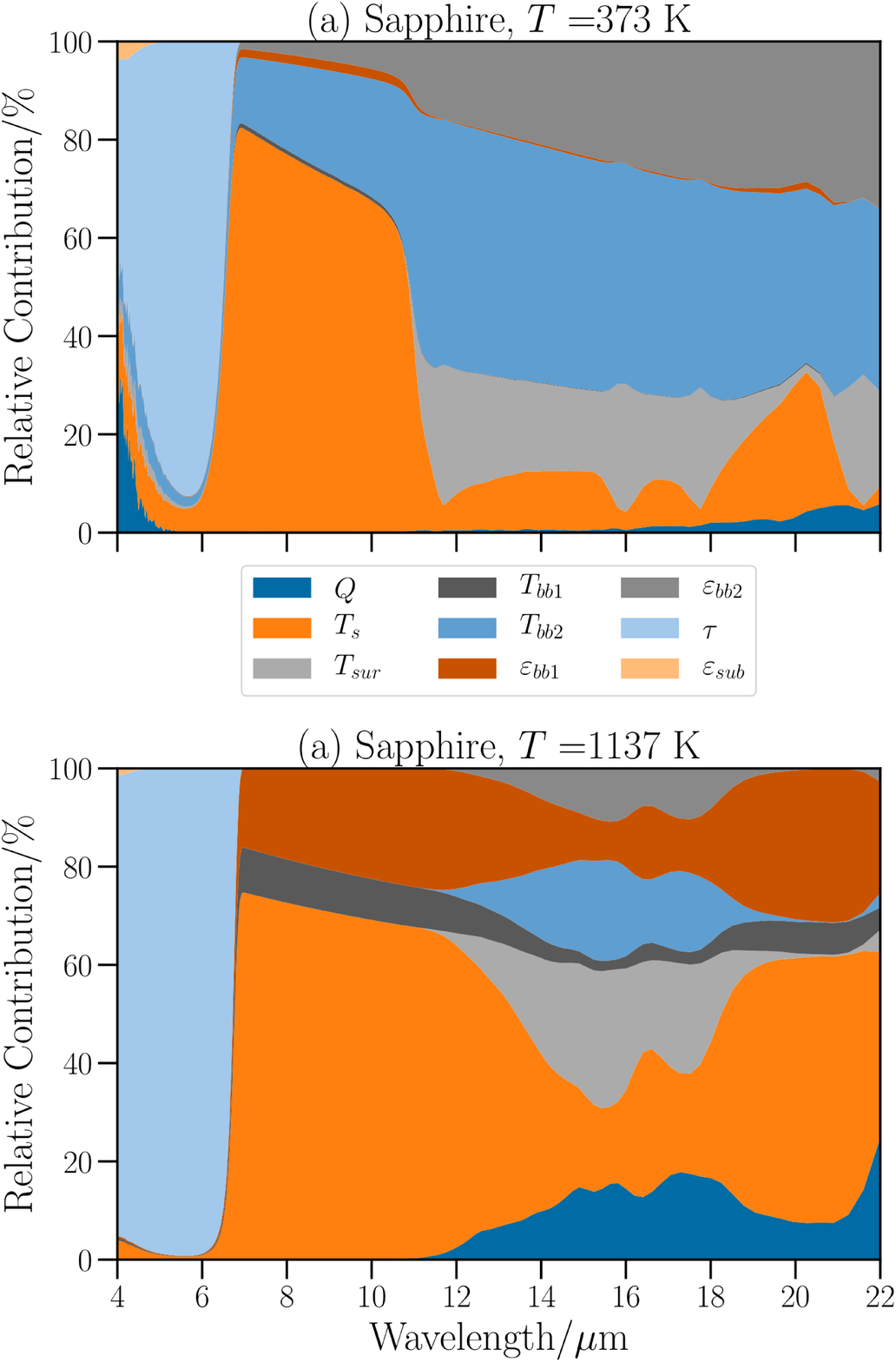}
\caption{Relative contribution of each term of Eq.~\ref{eq:final_uncertainty} to the total variance for sapphire at the lowest (a) and highest (b) temperatures measured. Note that the spectral range is restricted to $4-22$ $\mu$m.}
\label{fig:Al2O3_relativas}
\end{figure}

Finally, the relative contributions of each uncertainty source for both materials to the variance ($u^2_c$ from Eq.~\ref{eq:final_uncertainty}) are shown in Figs.~\ref{fig:Ni_relativas} and \ref{fig:Al2O3_relativas}. Some commonalities can be appreciated. For example, for both materials there is a general increase with temperature of the relative importance of the uncertainties corresponding to the sample temperature ($T_s$) and the high-temperature source temperature and emissivity uncertainties ($T_{bb1}$, $\varepsilon_{bb1}$), whereas those corresponding to the low-temperature source ($T_{bb2}$, $\varepsilon_{bb2}$) and the surroundings temperature ($T_{sur}$) decrease. This is expected, as these latter sources of uncertainty correspond to low-temperature radiance sources, the influence of which is greatly reduced when the emission from the sample is increased at higher temperatures. Nevertheless, some influence of them can still be observed at long wavelengths for the highest temperatures, especially for Ni.

Some differences in the relative weights for each material can also be distinguished. The weight corresponding to the radiance uncertainty ($Q$) is greater for Ni, due to the low signal reaching the detector from this low-emissivity sample. This contribution becomes important at higher temperatures (where most of the others are significantly reduced), or at short wavelengths and low temperatures (where the emitted radiation has a very low signal-to-noise ratio). On the contrary, this uncertainty is larger for sapphire at longer wavelengths. Finally, in the case of semi-transparent regions, the uncertainty due to the transmittance ($\tau$) is the dominant term, with the one corresponding to the emissivity of the substrate ($\varepsilon_{sub}$) being almost negligible. Overall, these Figures show similar trends as the ones shown for the previous uncertainty budget \cite{doi:10.1063/1.3431541}, but with greater detail and rigour. % Discutir el budget anterior con detalle.

%--------------------------------------------

\section{Calculation and propagation of uncertainties for integrated quantities }
\label{sec:monte_carlo}

\subsection{Integration of spectral data}

The sections above have dealt with measurements of the spectral directional emissivity. This parameter depends not only on temperature, but also on wavelength and emission angle; whereas, for most heat transfer applications, details on such dependences are often irrelevant. It is for this reason that integrated quantities, such as the total normal or total hemispherical emissivities, are often more useful for engineering applications. These quantities can be measured by radiometric or calorimetric methods, which rely on very different approaches but have been shown to be equivalent \cite{Monchau2018}. For radiometric methods, such as the one described in this work, the total directional emissivity (of which the total normal is a special case) and the total hemispherical emissivity are defined from their spectral counterparts as follows \cite{howell2015thermal}:

\begin{equation}
\varepsilon_T(\theta,T) = \frac{\int_0^{\infty}\varepsilon(\lambda,\theta,T)L(\lambda,T)d\lambda}{\int_0^{\infty}L(\lambda,T)d\lambda}
\end{equation}

\begin{equation}
\begin{split}
\varepsilon_H (T) &= \frac{1}{\pi}\int_0^{2\pi}d\phi\int_0^{\pi/2}\varepsilon_T(\theta,T) \cos(\theta)\sin(\theta)d\theta  \\
&\equiv \int_0^{\pi/2}  \varepsilon_T(\theta,T)\sin(2\theta)d\theta
\end{split}
\label{eq:hemisferica}
\end{equation}
where $\varepsilon_T(\theta)$ is the total directional emissivity at angle $\theta$, $\varepsilon_H$ is the total hemispherical emissivity, $\theta$ is the polar angle and $\phi$ is the azimuthal angle, which is assumed to bear no influence \cite{howell2015thermal}.

A common problem of radiometric methods is the difficulty of measuring the entire spectral range in which thermal radiation may be emitted (i.e., from the far-infrared to the visible range). A method for calculating the total normal emissivity of metallic materials from directional spectral measurements in a restricted spectral range was introduced in Ref.~\cite{SETIENFERNANDEZ2013390}. This method involved calculating the total normal emissivity through a numerical integration of the available spectral data and two possible extrapolations, an overestimation and an underestimation. Therefore, the best estimate was found to be the average of the values obtained using each of the two extrapolation procedures, with the expanded uncertainty being half the difference between them. Based on the general knowledge that the wavelength dependence of the emissivity of metals is monotonic throughout the infrared range \cite{howell2015thermal}, this extrapolation procedures assumes that the possible values are contained within the last measured data point at each end and the physical limit of the emissivity (either 0 or 1).

This method approaches the correct value of the total emissivity given a sufficiently wide spectral range without requiring actual information of the emissivity in the extrapolated region. This can be regarded as an application of the principle of maximum entropy, as described in the GUM \cite{gum}.

The maximum-entropy probability distribution for extrapolations of the spectral emissivity to the full electromagnetic spectrum when the available data is restricted to a spectral range [${\lambda_1},{\lambda_2}$], and the emissivity is known to be monotonically decreasing with wavelength, is given by:

\begin{equation}
\varepsilon_T = \frac{\pi}{\sigma T^4} \bigg [ \int_0^{\lambda_1} \varepsilon_1 L d\lambda +\int_{\lambda_1}^{\lambda_2} \varepsilon({\lambda})Ld\lambda +\int_{\lambda_2}^{\infty} \varepsilon_2 L d\lambda \bigg ]
\label{eq:definition}
\end{equation}
where $\sigma T^4/\pi$ is the total radiance per steradian emitted by a blackbody and the $\varepsilon^{(i)}$ are random variables decribed by uniform probability density functions (PDFs) for each side of the spectrum. In the particular case of metals, this becomes:

\begin{equation}
f(\varepsilon_i) = \bigg\{\begin{array}{lr} U(\varepsilon_{max},1) \text{ for } \varepsilon_1\\ U(0,\varepsilon_{min}) \text{ for } \varepsilon_2\end{array}
\label{eq:PDF}
\end{equation}
where $U(a,b)$ stands for the uniform distribution.

\begin{figure}[hptb]
\centering
\includegraphics[width = \linewidth]{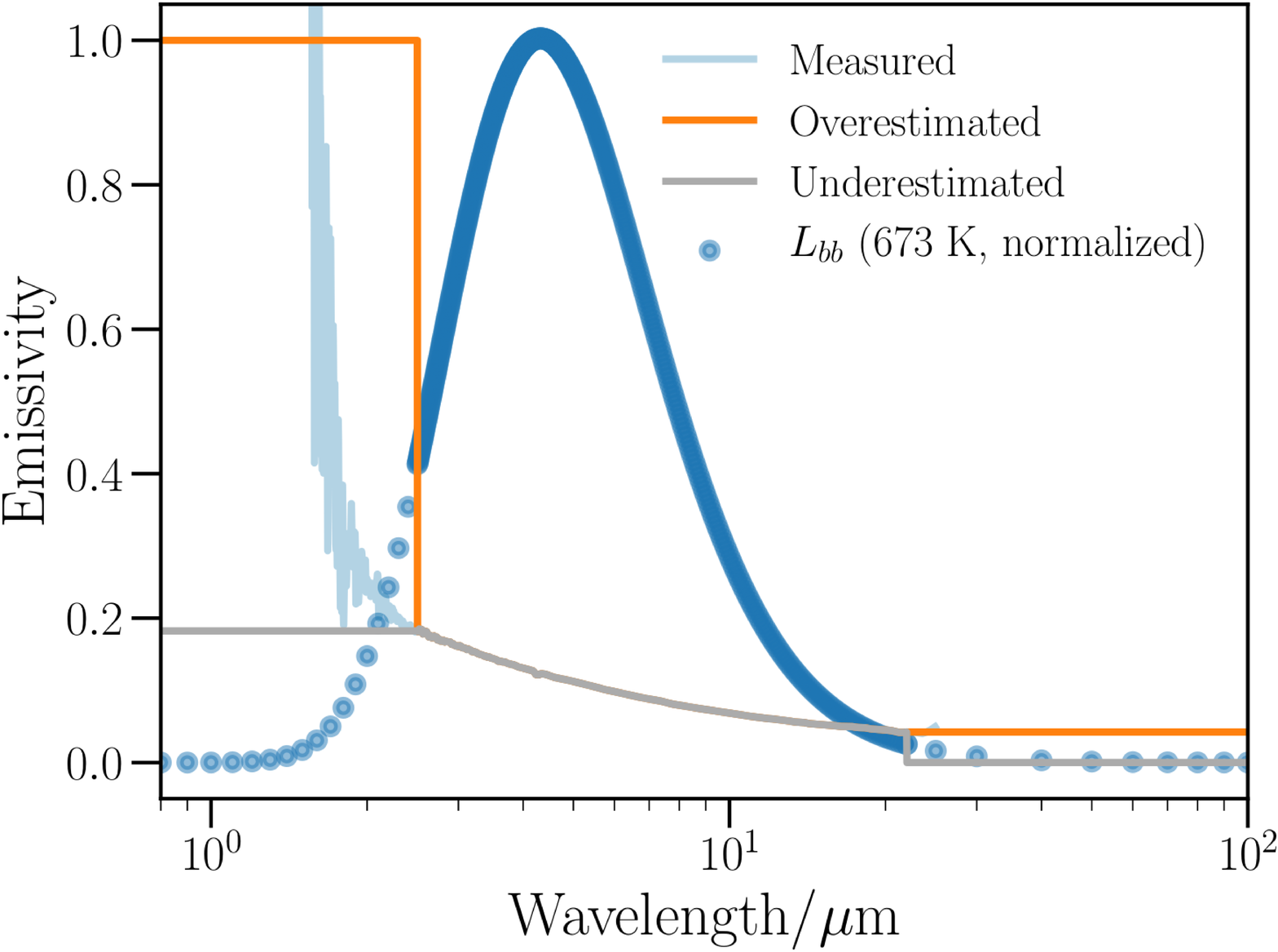}
\caption{Graphic representation of the extrapolation procedure for Ni using normal spectral data from Fig~\ref{fig:Ni_sombra}. Emissivity curves are shown as solid lines, corresponding to measured data (shaded blue), the overestimated limits (solid orange) and the underestimated ones (solid grey). A normalized blackbody radiation spectrum (blue dots) is shown to illustrate the relative weight of each spectral region.}
\label{fig:extrapolation}
\end{figure}

The final equation then becomes:

\begin{equation}
\varepsilon_T =  \varepsilon_1 F_{0 \rightarrow \lambda_1}(T) + \varepsilon_2 F_{\lambda_2 \rightarrow\infty}(T) + \frac{\pi}{\sigma T^4}\int_{\lambda_1}^{\lambda_2} \varepsilon_{\lambda}L_{\lambda}d\lambda
\label{eq:final}
\end{equation}
where $\varepsilon_1$ and $\varepsilon_2$ are uniform PDFs given by Eq.~\ref{eq:PDF}, $\lambda_1$ and $\lambda_2$ are the shortest and the longest wavelength of the experimental spectral interval, and $F_{a\rightarrow b}(T)$ stands for the fraction of the total radiance emitted by a blackbody in the $a-b$ wavelength range at temperature $T$. This function can be computed by numerical integration, series expansion \cite{howell2015thermal}, or analytic solutions based on polylogarithmic functions \cite{STEWART2012232}. An illustration of this method is shown in Fig.~\ref{fig:extrapolation} for normal spectral data of Ni at 673 K.

Eq.~\ref{eq:definition} and Eq.~\ref{eq:final} are based on the least-informative PDF given the only information that the emissivity in the chosen material is monotonic (as recommended by GUM in the case of lack of information). This is guaranteed in the particular case of metals due to the Drude law holding qualitatively for most of the range up to the visible range \cite{howell2015thermal}, but can be easily generalizable to any material with a known monotonic emissivity. Other types of extrapolations, such as assuming a constant emissivity value, have been used for the same purpose \cite{Monchau2018}. It must be noted that the approach followed in this work is conservative and might lead to larger uncertainty values, and other extrapolation procedures are possible for each case. Furthermore, since the weight of the Planck function decreases exponentially away from the Wien peak, the emissivity is only required to be monotonic for the spectral range in which the Planck function is non-negligible.

%Since many spectra have clearly defined tendencies away from the mid-infrared (e.g., ), it is possible to define extrapolation PDFs that follow these general tendencies, but which do not introduce biases induced by spurious data and extrapolations.

In the case of angular integration to obtain the total hemispherical emissivity (Eq.~\ref{eq:hemisferica}), the extrapolations are much simpler. The emissivity values at the extremes are given by the electromagnetic theory as $\varepsilon(0^{\circ})\simeq \varepsilon(10^{\circ})$ and $\varepsilon(90^{\circ})=0$ \cite{howell2015thermal}. This makes the calculation of the total hemispherical emissivity a simple generalization of the total normal one and with an uncertainty almost equal to those of the individual total directional emissivities.

\subsection{Propagation of uncertainties by a Monte Carlo method}

The complexity of Eq.~\ref{eq:final} makes the standard approach of propagation of uncertainties from spectral to total data difficult. In particular, the presence of an integral and the multiple instances of non-linear temperature dependence are the most demanding tasks. Therefore, the standard linearized approach to calculate the uncertainties is discouraged. A Monte Carlo method has been applied following the guidelines of the Supplement 1 to the GUM \cite{gum_montecarlo}. The propagation of uncertainties using Monte Carlo methods is common for complex models involving numerical integrals in the field of radiometry \cite{Esward_2007, Cordero_2007}. This approach is more general and less biased than the conventional method because it propagates the entire distribution functions describing each of the variables.

\begin{table}[hbtp]
\centering
\caption{Uncertainty sources for the calculation of the total emissivities and their distributions.}
\begin{tabular}{lll}
Source & Symbol & Distribution \\
\hline
%Sample temperature & $T$ & Gaussian \\
Measured emissivities & $\varepsilon(\lambda_i)$ & Multivariate Gaussian \\
Extrapolated emissivities & $\varepsilon_1$, $\varepsilon_2$ & Rectangular 
\end{tabular}
\end{table}

The Monte Carlo method of propagation of uncertainties requires defining the input PDFs for each of the variables that feature in Eq.~\ref{eq:final} ($\varepsilon_1$, $\varepsilon_2$, and each of the $\varepsilon(\lambda_i)$ datapoints). The uncertainty in the measured sample surface temperature is not considered as an input parameter in this approach. The extrapolated emissivities are defined as rectangular PDFs (following Eq.~\ref{eq:PDF}), whereas the discrete spectral emissivity data points are drawn from a multivariate Gaussian distribution and assumed to be perfectly correlated ($r=1$). Quantification of the actual correlation is complicated, but assuming a perfect correlation gives the highest possible uncertainty, whereas assuming no correlation gives an unrealistically low uncertainty value. Therefore, $r=1$ for all $\varepsilon(\lambda_i)$ is regarded as a conservative but realistic estimate. This approach is similar to that followed in Ref.~\cite{Monte_2010} for the same type of calculation, but without a Monte Carlo approach. It calculates the difference between the maximum possible value within the standard uncertainty and the best estimate of the total emissivity, and assigns it as the standard uncertainty of this parameter. This estimation relies implicitly on an assumption of perfect correlation between all data points. In any case, a full Monte Carlo analysis is deemed more robust because of the complexity of Eq.~\ref{eq:final}. In particular, the GUM Supplement warns of significant errors in the determination of both the mean and the standard uncertainty when performing these calculations for heavily non-linear functions \cite{gum_montecarlo}.

In this work, a Monte Carlo method has been implemented in Python 3.7. Standard functions from the NumPy (v1.16.4) and SciPy (v1.3.0) libraries have been used. The pseudo-random generator used has been the standard Mersenne Twister (\texttt{MT19937}) implemented in that version of NumPy. Paralellization issues have been avoided by running the program serially \cite{Esward_2007}. Two different sets with numbers of trials of $M=100$ and $M=2\cdot10^5$ have been calculated from an initial high-entropy seed, while different seed values have also been tried to check the independence of the results from numerical biases. The latter number of trials has been selected as appropriate for a 95\% confidence interval (coverage factor $k=2$), according to the recommendations of Ref.~\cite{gum_montecarlo}, whereas the former has been used as a check to verify that the statistical properties of the result are stable and do not evolve significantly with the number of trials. %Other sample statistics may require a larger number of trials.

\begin{figure}[hptb]
\centering
\includegraphics[width = \linewidth]{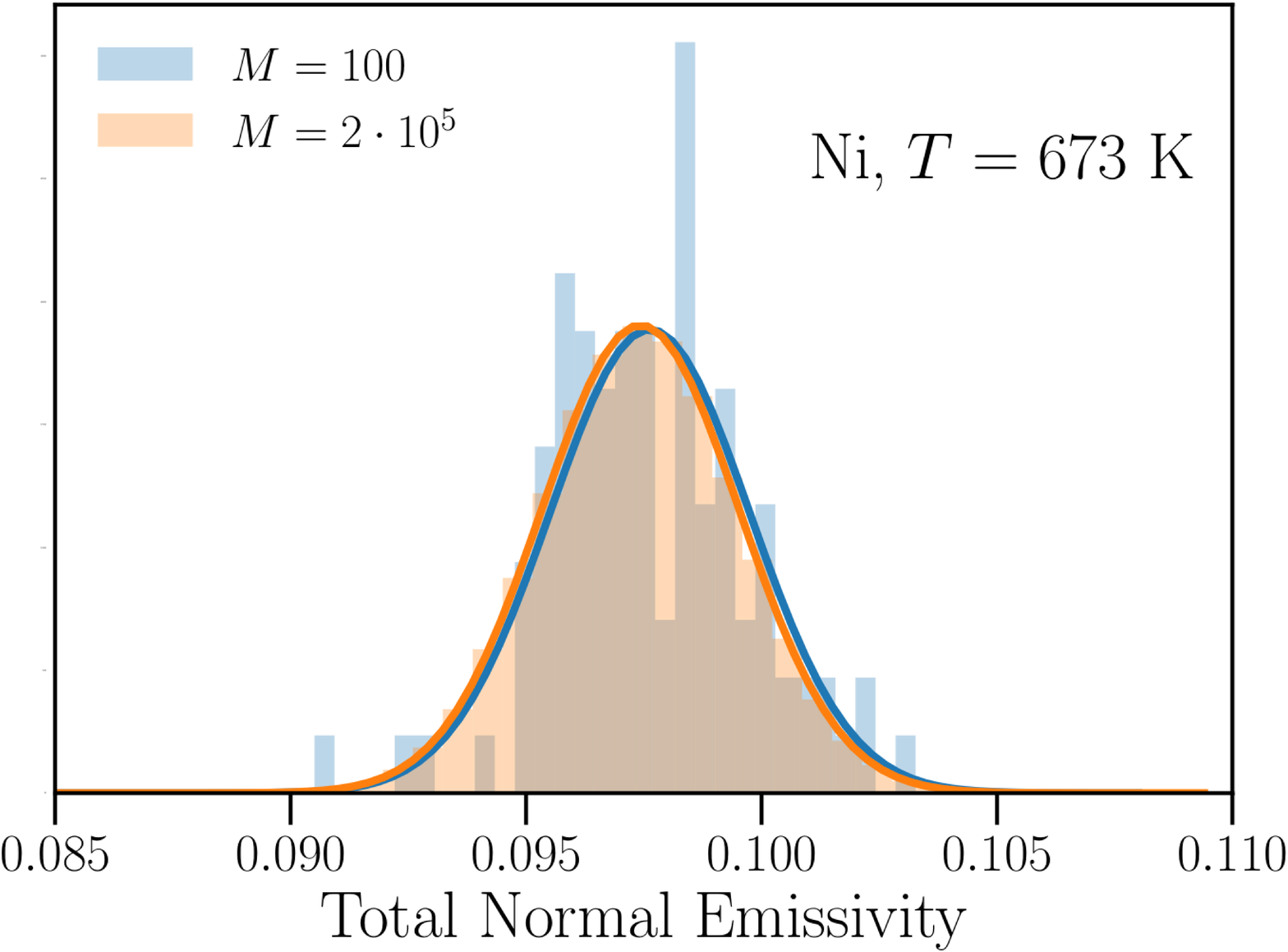}
\caption{Calculation of the total normal emissivity of Ni at 673 K and propagation of its uncertainty by a Monte Carlo method using two different numbers of trials ($M=100$ and $M=2 \cdot 10^5$). Solid lines correspond to Gaussian fittings to the histograms.}
\label{fig:histograms}
\end{figure}

\begin{figure}[hptb]
\centering
\includegraphics[width = \linewidth]{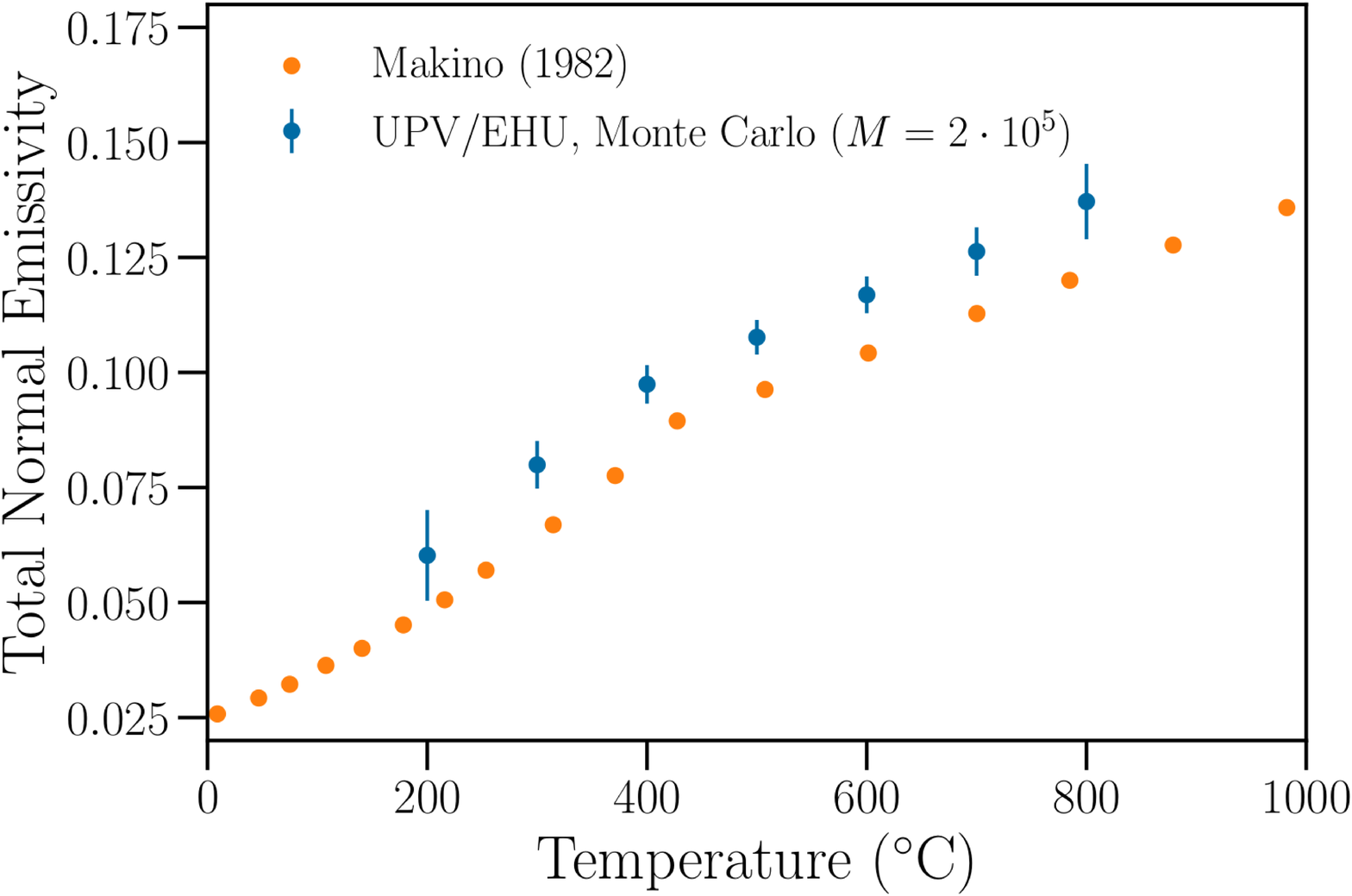}
\caption{Total normal emissivity and its standard uncertainty ($k=2$) of Ni calculated using the Monte Carlo method for $M=2 \cdot 10^5$. Literature data from Ref.~\cite{makino1982study} are shown for comparison.}
\label{fig:total_Ni}
\end{figure}

The resulting PDFs from the calculations are shown in Fig.~\ref{fig:histograms} for one temperature, together with Gaussian fittings to the data. It can be seen that both means and standard deviations agree satisfactorily, even though the histograms are very different. The results of these calculations for all temperatures are shown in Fig.~\ref{fig:total_Ni}. Literature data from Ref.~\cite{makino1982study}, obtained with a similar radiometric method, is shown for comparison. It can be seen that the results obtained in this work compare favourably to those reported in the literature in a semi-quantitative manner, although a clear offset is appreciable. Further work needs to be done to clarify this point and advance on the development of reliable metallic reference materials.

\begin{figure}[hptb]
\centering
\includegraphics[width = \linewidth]{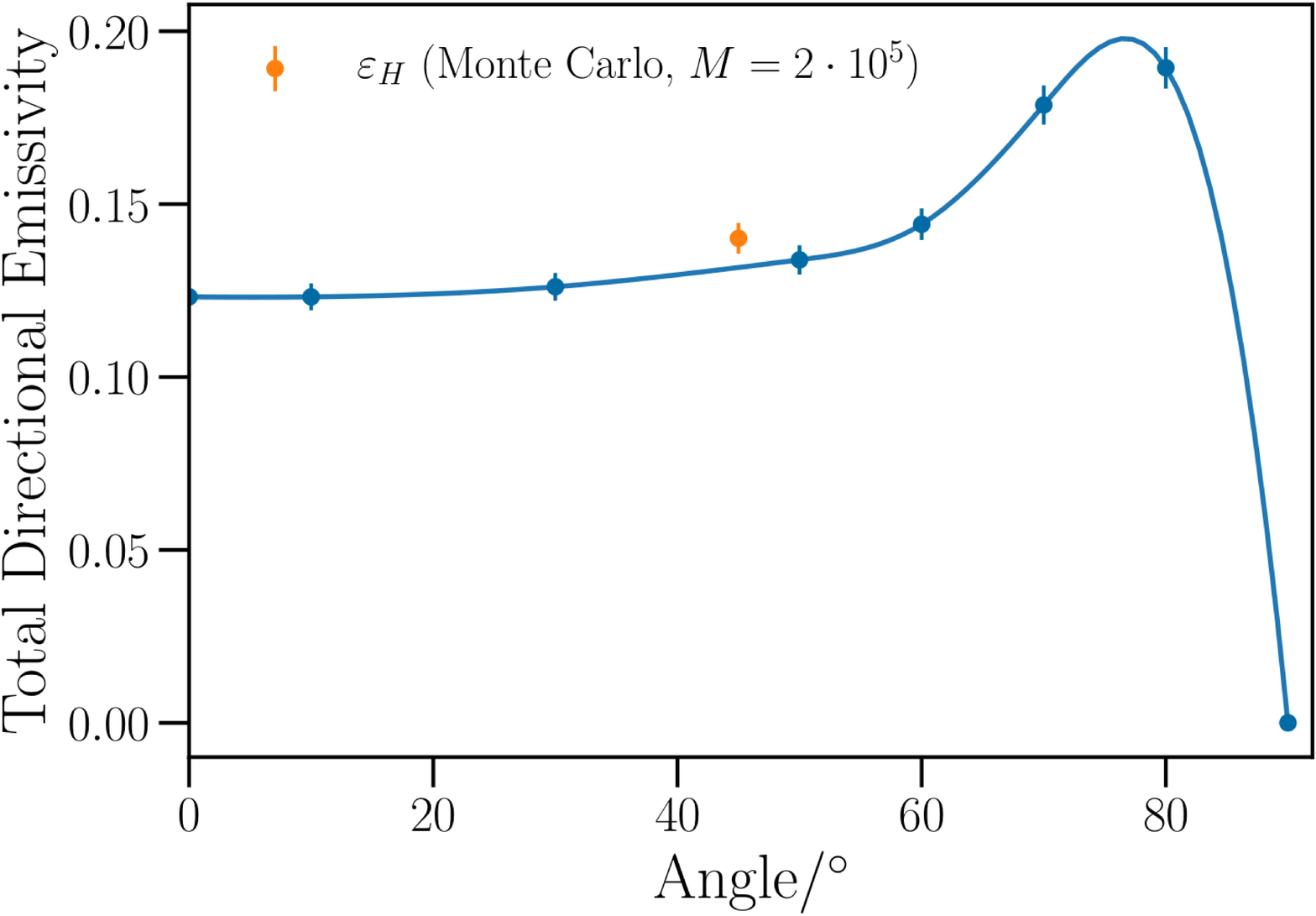}
\caption{Total directional emissivities and combined expanded uncertainties ($k=2$) of a V-4Cr-4Ti alloy at 673 K (blue dots), and total hemispherical emissivity computed using a Monte Carlo method (orange dot). The solid line is a calculated spline which is used for the integration of the data.}
\label{fig:vanadium_directional}
\end{figure}

\begin{figure}[hptb]
\centering
\includegraphics[width = \linewidth]{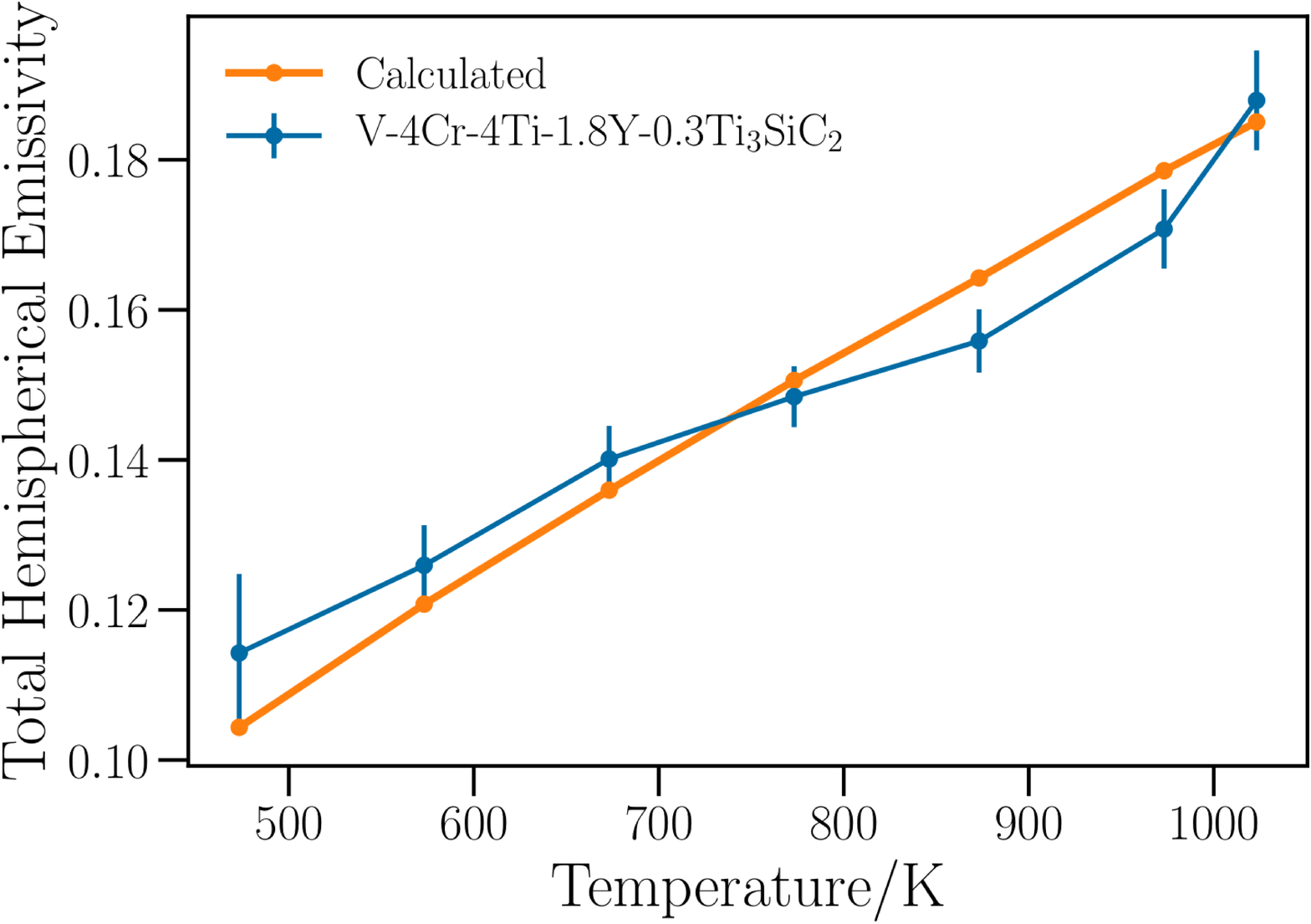}
\caption{Total hemispherical emissivities and combined expanded uncertainties ($k=2$) of a V-4Cr-4Ti alloy as a function of temperature \cite{ECHANIZ201986}. The expected tendency, given by the basic free-electron theory, is shown for comparison.}
\label{fig:vanadium_hemispherical}
\end{figure}

In order to illustrate this approach also for the integration of directional data, results corresponding to a recently published study on V-4Cr-4Ti alloys have been used \cite{ECHANIZ201986}. These materials are regarded as candidate structural materials for nuclear fusion reactors, and so their total radiative power is of key interest. Directional spectral data are integrated to yield the total directional emissivities and the total hemispherical one. Fig.~\ref{fig:vanadium_directional} shows the total directional emissivities with their uncertainties and the calculated total hemispherical emissivity at 673 K. A multivariate Gaussian distribution with complete correlation ($r=1$) has been assumed again as a conservative estimate. The resulting values for the temperature-dependent total hemispherical emissivity of this alloy is shown in Fig.~\ref{fig:vanadium_hemispherical}. The results show a moderate agreement to the values predicted by the free-electron theory from literature electrical resistivity data, as reported in Ref.~\cite{ECHANIZ201986}. Discrepancies can be attributed to the non-ideal nature of the measured surface, and to the simplicity of the Hagen-Rubens theory that relates the emissivity to the electrical properties of materials \cite{howell2015thermal}. It should also be noted that the theoretical estimate also relies on the quality of the electrical resistivity data used, for which no reported uncertainty or detailed microstructure are available.

\section{Conclusions}

Emissivity measurements are complex procedures that require robust experimental methods. This work details an update of the framework followed at the UPV/EHU for the emissivity measurements performed with the HAIRL emissometer, with particular emphasis in the experimental and methodological improvements. 

Concerning the former, the temperature and vacuum ranges have been broadened significantly. With respect to the latter, there are three main modifications applied to the measurement equation. First, the inclusion of optical path differences and reference emissivities; second, the reformulation of the measurement equation to avoid correlation between the calibration parameters; and third, the addition of different temperature measurement methods for metals and ceramics. Furthermore, a systematic review of the uncertainty budget following the ISO GUM guidelines has been applied to the spectral data. Finally, an integration and extrapolation procedure for calculation of total emissivities, and their uncertainty calculations using a Monte Carlo method are also reported.

This framework has been applied successfully to metallic and ceramic materials, and is expected to be very useful to current and future development of similar emissivity-measuring devices.

\section*{Appendix A. Numerical uncertainty budgets}

\begin{table}[hpbt]
\caption{Uncertainty budget for Ni at 673 K and 5 $\mu$m.}
\begin{center}
\begin{tabular}{lll}
Source of uncertainty & Type & Value (\%) \\
\hline
Sample signal repeatability & A & 0.07 \\
High-$T$ blackbody signal repeatability & A & 0.08 \\
Low-$T$ blackbody signal repeatability &  A & 2.24 \\
Optical path anisotropy &  A & 0.08 \\
Sample temperature &  & 0.27 \\
Surroundings temperature &  & 0.43 \\
High-$T$ blackbody temperature  &  & 0.10 \\
Low-$T$ blackbody temperature &  & 0.44 \\
High-$T$ blackbody emissivity & B & 0.29 \\
Nextel 811-21 emissivity & B & 1.03 \\
\hline
Relative combined standard uncertainty & & 1.55\\
\hline
Relative expanded uncertainty ($k=2$) & & 3.10\\
\end{tabular}
\end{center}
\label{table:budget_ni}
\end{table}

\begin{table}[hpbt]
\caption{Uncertainty budget for $\alpha$-Al$_2$O$_3$ at 373 K and 8 $\mu$m.}
\begin{center}
\begin{tabular}{lll}
Source of uncertainty & Type & Value (\%) \\
\hline
Sample signal repeatability & A & 0.36 \\
High-$T$ blackbody signal repeatability & A & 0.08 \\
Low-$T$ blackbody signal repeatability &  A & 0.40 \\
Optical path anisotropy &  A & 0.10 \\
Sample temperature (Christiansen) &  & 0.29 \\
Surroundings temperature  &  & 0.43 \\
High-$T$ blackbody temperature &  & 0.10 \\
Low-$T$ blackbody temperature & & 0.44 \\
High-$T$ blackbody emissivity & B & 0.29 \\
Nextel 811-21 emissivity & B & 1.03 \\
\hline
Relative combined standard uncertainty & & 2.09\\
\hline
Relative expanded uncertainty ($k=2$) & & 4.18\\
\end{tabular}
\end{center}
\label{table:budget_al2o3}
\end{table}

\newpage

\section*{Acknowledgements}

I. Gonz\'alez de Arrieta acknowledges the Basque Government for its support through a PhD fellowship.

%\section*{References}

%\bibliography{bibfile}

\end{document}